\documentclass[aps,preprint]{revtex4-1}

\usepackage{comment} 
\usepackage{graphicx}
\usepackage{amsmath}
\usepackage{setspace}  
\usepackage{amssymb} 
\usepackage{pstricks}
\usepackage{tikz}
\usepackage{multirow}
\usepackage{float}
\usepackage{soul}

\usepackage[latin1]{inputenc}
\usepackage{hyperref}
\usepackage{colortbl}
\usepackage{tikz}
\usetikzlibrary{shapes,arrows}
\usepackage{geometry}
\usepackage{hyperref}
\usepackage[T1]{fontenc}

\newcommand{\be}{\begin{equation}}
\newcommand{\ee}{\end{equation}}

\begin{document}

%\maketitle
%\begin{minipage}[h]{\textwidth}
\title{Mesoscale simulations of shockwave energy dissipation via chemical reactions}
\author{Edwin Antillon, Alejandro Strachan}
\email{Corresponding author: strachan@purdue.edu}
\affiliation{School of Materials Engineering and Birck Nanotechnology Center Purdue University, West Lafayette, Indiana, 47907, USA}
\date{Feb,1 2015}					% used by \maketitle

\maketitle

We use a particle-based mesoscale model that incorporates chemical reactions at a coarse-grained level to 
study the response of materials that undergo volume-reducing chemical reactions 
under shockwave-loading conditions. We find that such chemical reactions can attenuate the shockwave and characterize 
how the parameters of the chemical model affect this behavior. The simulations show that the magnitude of the volume collapse 
and velocity at which the chemistry propagates are critical to weaken the shock, whereas the energetics in the reactions play only 
a minor role. Shock loading results in transient states where the material is away from local equilibrium and, interestingly, chemical
reactions can nucleate under such non-equilibrium states. Thus, the timescales for equilibration between the various degrees of
freedom in the material affect the shock-induced chemistry and its ability to attenuate the propagating shock.

\section{Introduction}

%{\it Describe goals, what is understood, what are the challenges}

When materials are subject to dynamical mechanical loads (shockwaves) a plethora of complex processes
are launched in response to the insult. These can include plastic deformation, \cite{HolianScience1998,BringaScience2005,LoveridgePRL2001}, 
phase transitions, \cite{KadauScience2002,KalandarPRL2005,Erskine1991}, chemical reactions \cite{YangJAP2004,StrachanPRL2003}, and even 
electronic transitions \cite{ReedNatPhys2008}.
Studying the response of different materials to shockwaves has resulted in significant insight into the response of 
materials under extreme conditions.\cite{DlottARPC2011} In most cases, when solid materials are shocked above a threshold strength, known as the Hugoniot
elastic limit, plastic deformation nucleates behind the shock front to release the compressive stress along the shock 
direction and minimize free energy by achieving a more hydrostatic state.  \cite{GermannPRL2000,JaramilloPRB2007}
This stress relaxation weakens the leading shockwave, often resulting in a two-wave structure, with a leading elastic wave
followed by a plastic wave that propagates at slower speeds; similar two-wave structures have been found in shock induced martensitic 
transformations \cite{KadauScience2002}. In the case of explosives, exothermic reactions that lead to low density 
products can enhance the initial shockwave and transform it into a detonation wave. For this to happen, the chemical 
reactions have to generate enough energy and pressure. However, thermodynamics alone is not sufficient; fast reaction kinetics are 
critical to significantly affect a traveling shockwave.

Materials that dissipate or absorb the energy in shockwaves causing the shock front to weaken as it propagates are of interest for protection 
against impacts, collisions and blasts. Impact resistant materials include fiber composites \cite{kevlar} and ceramic/metal armor \cite{lopez2005effect}. 
There has been growing interest in softer materials capable of absorbing shockwave energy via molecular-level processes.
Polyurea, a polymer with glass transition below room temperature, has been shown effective at absorbing energy in ballistic impact 
tests. \cite{bogoslovov2007impact} While the mechanisms for the high energy absorption capability at high strain rates are not fully understood, a 
transition to the glassy state is believed to play a key role. \cite{bogoslovov2007impact,grujicic2010computational,grujicic2013coarse} 
Materials with very high porosity at the nanoscale, such as metal organic frameworks, are also attracting attention for such applications 
as void collapse can contribute to weaken the shockwave \cite{Yot2014}.

In this paper, we explore the possibility of using chemical reactions in molecular systems for shock wave energy absorption; specifically
endothermic, volume reducing chemistry. We use a recently developed model for coarse grain simulations for a class of model materials
that exhibit the desired behavior at the molecular level and and study the mechanisms through which shock energy dissipation can occur.
Our objective is to understand requirements of such shockwave energy dissipating (SWED) materials and to establish how the characteristics
of the chemical reactions affect its ability to weaken shocks. We foresee this knowledge could contribute to experimental design efforts.
Our simulations demonstrate that a chemical reaction front involving endothermic, volume-reducing chemistry 
can propagate fast enough to couple with the leading shockwave and weaken it. The results also
shed light on how the nucleation and propagation of such chemical reactions occur under dynamical loading.

The paper is organized as follows. Section II introduces ChemDID, the coarse grain model used to investigate shock-induced chemistry and the
family of model materials to be characterized. Section III discusses the response of materials that can undergo volume reducing chemistry
to shock loading and relate characteristics of the chemical reactions and the ability of the materials to weaken shocks. Section IV discusses
the molecular processes that govern the nucleation of chemistry behind the shockwave and Section V draws conclusions.

\section{Model and simulation details}

\subsection{ChemDID}

ChemDID \cite{Antillon2014} is a coarse-grained, particle-based model that enables the description of stress-induced chemical reactions involving degrees 
of freedom (DoF) internal to the mesoparticles. Unlike all-atom MD, where particles represent atoms, in ChemDID particles describe 
 extended objects, molecules in this paper. Such molecules will be represented by a spheres with 3N-3 internal DoFs (for an N atom molecule), out of which we will single out one of them to describe the chemical reactions in the molecule while the rest will be treated via the equipartition theorem.

\begin{figure}[h!]
  \centering
  %trims (crops) from left, bottom, right and top respectively
  \includegraphics[scale=.26]{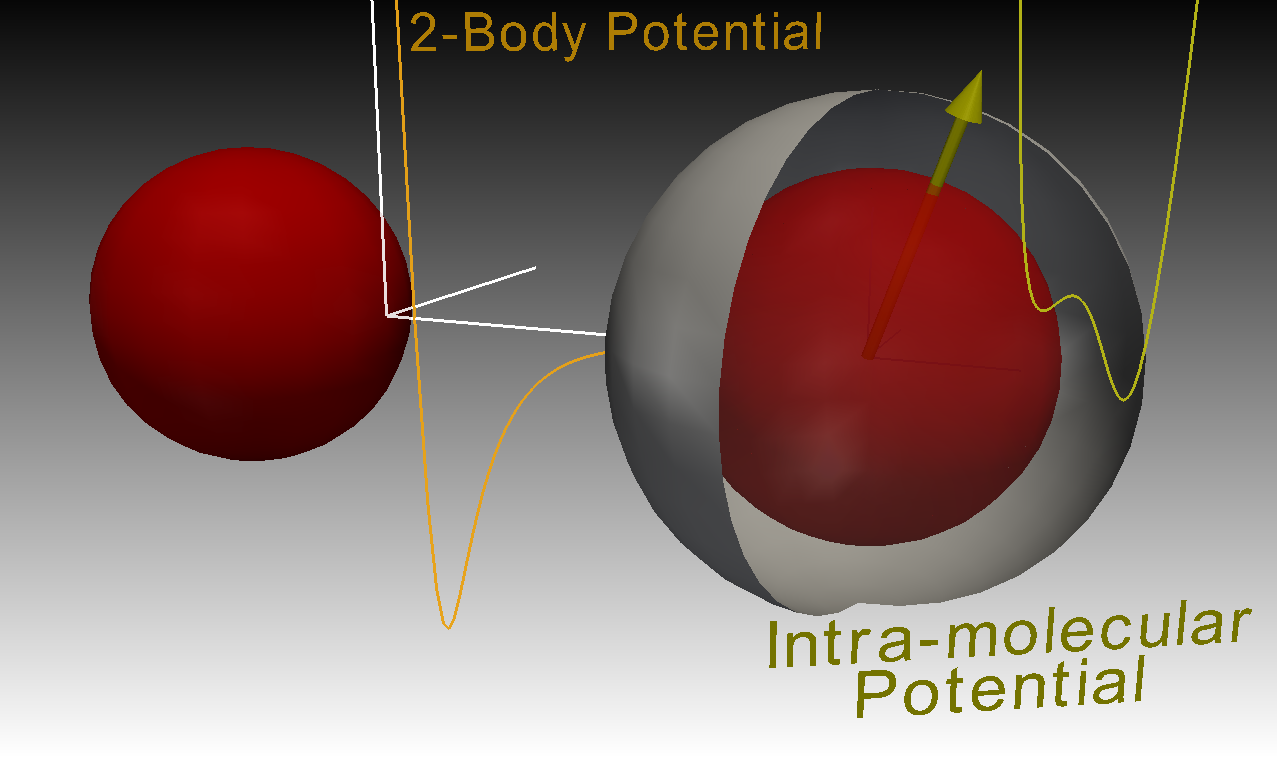}
  \caption{Depiction of intermolecular and intramolecular potentials in ChemDID}
  \label{fig:ChemDID}
\end{figure}

\begin{figure}[h!]
  \centering
  \begin{tabular}{cc}
    \includegraphics[trim=1cm 4cm 6cm 2cm,clip=true,scale=1.15]{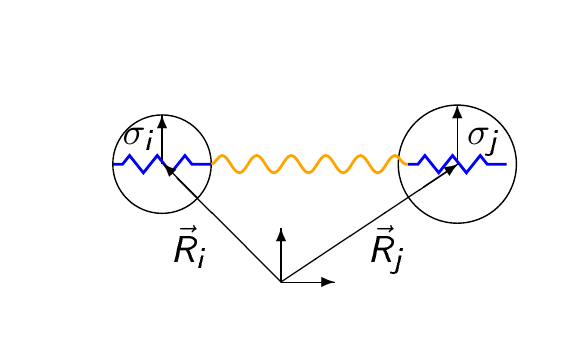} &
    \includegraphics[trim=0.0cm 1cm 3cm 1cm,clip=true,scale=.85]{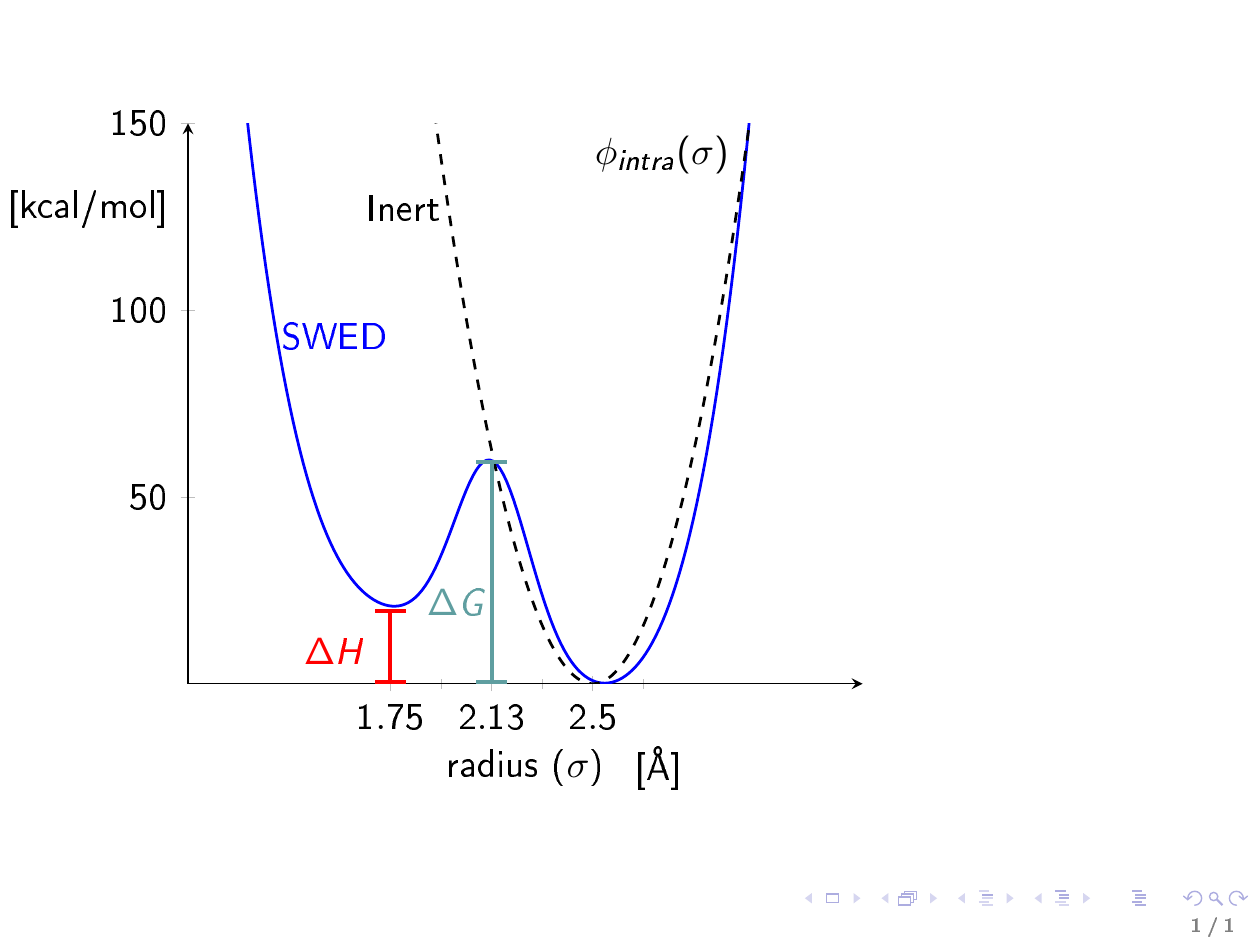} \\
    (a)  &  (b) \\
  \end{tabular}
  \caption{(a) Representation of ChemDID as a coupled spring system with an intermolecular and intramolecular terms.
    (b) The Intra-molecular potential used describes an Inert case (dashed line) and SWED case (solid line); The parameters $\Delta G$ determine the activation barrier energy and $\Delta H$ describes the amount of energy absorbed during a collapse to a low-volume state.}
  \label{fig:springs}
\end{figure}

The variable singled out will represents the molecular radius ($\sigma$), and is described with explicit Hamiltonian dynamics 
and an associated potential energy that can enable chemical reactions. The remaining 3N-4 DoFs are described in an averaged 
form using the approach proposed in Ref. \cite{StrachanPRL2005}. Their state is described by a single dynamical variable that represent their
temperature. Thus, the complexity of many-body intra-molecular interactions, including the desired volume collapsing reactions, is reduced to 
an Intra-molecular potential inspired in transition state theory, where the size of the mesoparticle is used as an order parameter 
that governs the transition between an initial high-volume low-energy state or a collapsed-low volume high-energy state; we will refer to such 
materials as SWED materials, while materials that do not undergo chemical reactions will be called {\it Inert}. 
Fig. \ref{fig:springs} contrasts the two cases described. The non-bonded interactions between mesoparticles are described by pair-potentials 
which acts from surface to surface distance as shown in Figure \ref{fig:ChemDID}.

The Hamiltonian of the system is:

\begin{align}\label{eqn:Hamiltonian}
  H(\{R_i\},\{\sigma_i\},\{P_i\},\{\pi_i\}) = & \sum_{i<j} \phi_{inter}( |\vec{R_i}-\vec{R_j}|-\sigma_i-\sigma_j) +  \sum_{i} \phi_{intra}( \sigma_i) \nonumber \\ 
  & + \sum_{i} \frac{P_i^2}{2 m_i} + \sum_{i} \frac{\pi_i^2}{2 m_i^{*}}
\end{align}
where $\phi_{inter}$ and $\phi_{intra}$ describe the inter-molecular and intra-molecular potential terms, $\vec{P_i}$ describes the translational 
momentum and associate mass $m$ of particle $i$, and $\pi_i$ similarly describes the conjugate momentum to the breathing mode with its 
associated inertial parameter $m^*_i$. A derivation of the equations of motions has been previously shown in \cite{Antillon2014}

We note that the Hamiltonian does not account for the remaining 3N-4 internal DoFs of the molecules. These modes exchange energy with the Hamiltonian variables
(center of mass position and molecular radius) and they are described with the approach proposed in Ref. \cite{StrachanPRL2005}. These internal DoFs are 
incorporated statistically and described by temperature ($T^{int}$); they are coupled to the explicit DoFs via the position update equation such that energy flows 
to equalize the temperatures associated with the various degree of freedom: $T^{int}$, that of the radial breathing mode $T^{rad}$ and that of the c.m. of the 
molecules $T^{mol}$. The resulting equation of motion for the internal temperature, see Refs. \cite{StrachanPRL2005,LinJCP2014}, is:

\begin{equation}\label{eqn:dTint}
  \frac{ \dot{E}^{int}_i }{C^{int}_i} 
  = \dot{T}^{int}_i =    \nu_{meso} \frac{(T^{meso}_i -T^{int}_i)}{m_i C^{int}_i \langle \omega_{inter}^2 \rangle \Theta_o} |\vec{F}^{inter}_i|^2 + 
  \nu_{rad} \frac{(T^{rad}_i -T^{int}_i)}{m^*_i C^{int}_i \langle \omega_{rad}^2 \rangle \Theta_o} |F^{rad}_i|^2
\end{equation}

where $E^{int}_i$ is the energy of the implicit DoFs, $C^{int}_i$ is their specific heat, $\nu_{meso}$ and $\nu_{rad}$ describe the strength of the 
internal-to-intermolecular coupling and the internal-to-radial coupling respectively,
 $\Theta_o$ is a reference temperature, and the ratio $|\vec{F}|^2/(m \langle \omega^2 \rangle$) provides a natural timescale for the corresponding 
 interaction. Note that the magnitude and direction of the energy flow between internal DoFs and the molecular centers of mass (first term in the
 RHS of Eq. \ref{eqn:dTint}) and the breathing mode (second term) is governed by the difference in local temperatures such that heat flows
 from hot to cold. The total energy is a conserved quantity in ChemDID and it is given by 
\begin{align}
 E_{tot} = & \sum_i \phi_{inter}(|R_i-R_j|-\sigma_i-\sigma_j)+ \sum_i \phi_{intra}(\sigma_i) \nonumber \\
 &  + \sum_i \underbrace{ \frac{P^2_i}{2 m_i}}_{\LARGE \frac{3}{2} k_B  T^{meso}_i} + \sum_i \underbrace{\frac{\pi_i^2}{2 m^*_i}}_{\LARGE \frac{1}{2} k_B T^{rad}_i} + \sum_i E^{int}_i  
\end{align}
We remind the reader that although the above definition of temperatures only applies for equilibrium conditions after time or ensemble averages in the
canonical ensemble, defining instantaneous and local values is useful to study processes out-of-equilibrium.
\subsection{A model reactive molecular crystal}

The parameterization of a ChemDID model involves determining inter-molecular and intra-molecular potentials, the molecular mass
and the inertial parameter for the dynamics of the molecule radii and as well as the coupling constants, $\nu_{meso}$ and  $\nu_{rad}$, 
that describe the coupling between the internal DoFs and the molecular centers of mass and radii. In this paper we parameterized a 
model reactive material with thermo-mechanical properties similar to anthracene, a molecular material believed to be capable of endothermic, 
volume reducing chemistry; DFT calculations by Slepetz et al. \cite{slepetz} show that the low-volume endothermic state of anthracene has an energy 
between 10-20 kcal/mol over the high-volume ground state.

We studied a family of intramolecular potential associated with the breathing mode in order to quantify how characteristics of
the chemical reactions affect the shock weakening power of the material. We characterize the chemical reactions by three
key parameters: i) volume collapse, ii) endothermicity, and iii) activation energy. The initial radius of the molecule is taken as 
$\sigma$  = 2.50 \AA, this value corresponds roughly to one of the dimensions that undergoes the greatest change in the 
anthracene molecule under the application of pressure\cite{slepetz,Jezowski}. The intra-molecular potential used takes the
form:

\begin{equation}\label{eqn:intrapot}
  \phi_{intra}( \sigma) = K*(\sigma-\sigma_{min})^2 \cdot(\sigma-\sigma_{max})^2+ \Delta H~\frac{(\sigma-\sigma_{max})}{(\sigma_{min}-\sigma_{max})} 
  + A~e^{-\frac{(\sigma-(\sigma_{min}+\sigma_{max})/2)^2}{2\sigma_o^2}}
\end{equation}
,where $\sigma_{min}$ and $\sigma_{max}$ denote the low-volume and high-volume (meta)stable points, the parameter $K$
 determines the overall curvature of the potential away from the (meta)stable points, while a Gaussian term (A,$\sigma_o$) controls the curvature of 
the region in-between, $\Delta H$ denotes the amount of endothermicity of the reaction for the collapsed state at $\sigma = \sigma_{min}$, and the
activation barrier $\Delta G$ will depend implicitly on both the Gaussian parameters (A,$\sigma_o$) and the overall curvature parameter $K$.
Table \ref{table:thetable} shows the intramolecular potential parameters.

In determining the overall collapse of the molecule, the radius of the sphere and the inter-molecular separation will both play a role.
In this paper, we  quantify this volume-collapse in terms of its van der Waals radius 
\begin{equation}\label{eqn:VvdW}
  V_{vdW} = \frac{4}{3} \pi (\underbrace{ \sigma + \Delta_{vdW}}_{r_{vdW}})^3
\end{equation}
where van der Waals radius ($r_{vdW}$) will be given by the sum of the sphere's hard-core radius ($\sigma$) and a van der Waals skin 
($\Delta_{vdW}$). The later depends on the intermolecular potential, discussed next, and takes the value to 1.07\AA. 
%Using this value for the lattice constant and the initial rigid radius we can get the skin distance to be: $\Delta_{vdW}$  = 1.07 \AA. 
A Morse potential will be used to describe its inter-molecular interactions, given by:
\begin{equation}\label{eqn:morse}
  \phi_{inter}(R_{ij}-\sigma_i-\sigma_j) =   \epsilon_0  [e^{\gamma        (1-(R_{ij}-\sigma_i-\sigma_j)/r_0)}  
                                                    - 2 e^{\frac{\gamma}{2}(1-(R_{ij}-\sigma_i-\sigma_j)/r_0)} ]
\end{equation} 
where the parameters $\epsilon_0$, $r_0$, and $\gamma$ describe the cohesive energy, interaction range, and its curvature near its global minima; the values used are shown in Table \ref{table:thetable}. The ground state structure of such a system is the fcc crystal structure and the inter- and intra-molecular parameters chosen resulting a lattice parameter of $a = 10.1$ \AA. The molecular mass will be taken as $m = 296.1$ g/mol, which has been used previously as a benchmark for molecular crystals and corresponds to an HMX-molecule \cite{JaramilloPRB2007,lynch2009coarse}. This gives an equilibrium volume per molecule ($V_{eq}$ ) of 258.8 \AA$^3$, close to that of athracene (231.2 \AA$^3$). The equilibrium volume per molecule and the van der Waals volumes are related by the packing fraction $(\equiv \frac{V_{vdW}}{V_{eq}})$, which corresponds to a value of 0.74 for an fcc arrangement.

\begin{figure}[h!]
  \centering
  \begin{tabular}{cc}
\hspace{-1.25cm}    \includegraphics[scale=0.22]{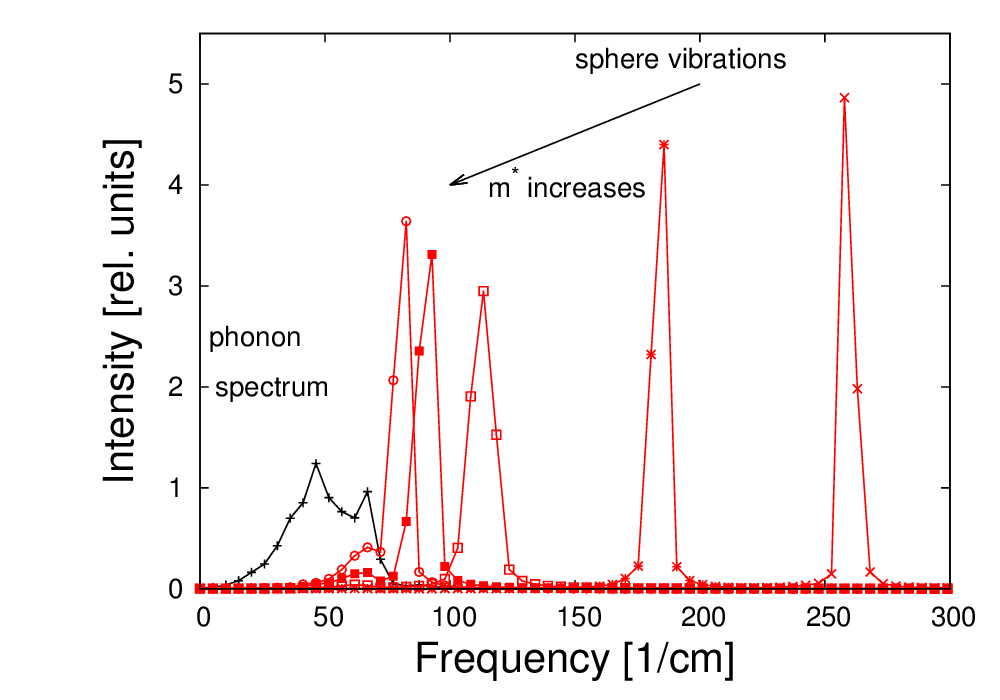} & 
    \includegraphics[scale=0.24]{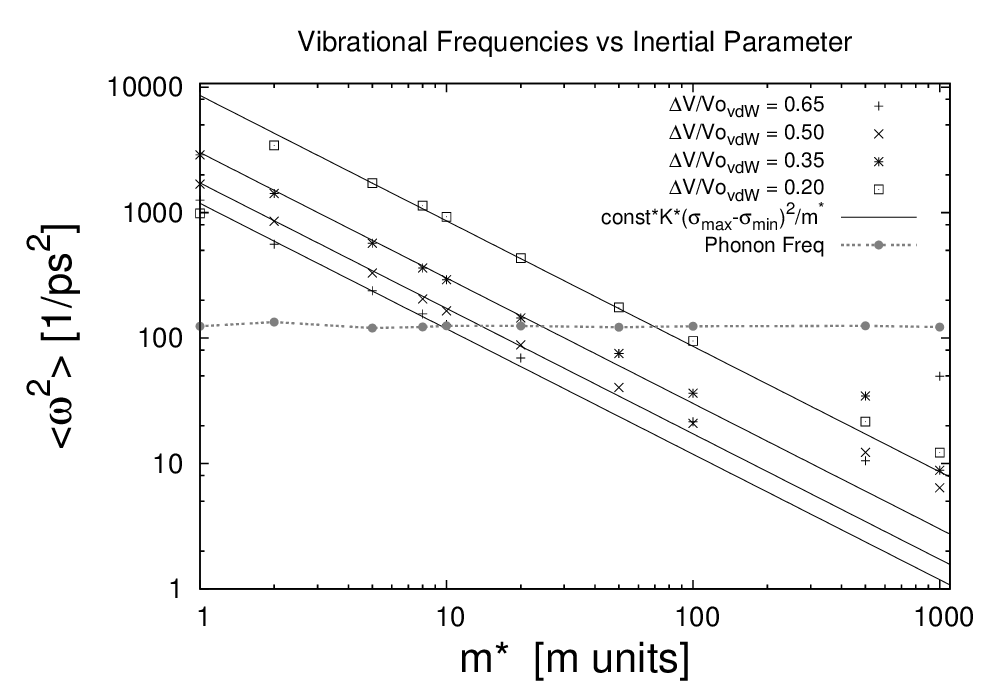}\\
    (a) & (b) \\
  \end{tabular}
  \caption{(a) Black and red curves correspond to the phonon and radial breathing modes - with different $m^*$- respectively vs frequency.
    (b) Dotted and solid lines correspond to the mean vibrational frequency squared  $\langle \omega^2 \rangle$ for the phonon and breathing modes (for different volume-collapse) respectively vs inertial parameter ($m^*$); the expected behavior $\langle \omega^2 \rangle \propto \frac{1}{m^*}$ can be observed, where the proportionality factor is given by the curvature of the intramolecular potential at its global minima.}
  \label{fig:DoS}
\end{figure}

The remaining parameters in the model determine the coupling constants between the various DoFs and to do this it
is useful to look at ChemDID the in terms of a coupled spring model, where a set of intra-molecular 
spring are in series with inter-molecular springs as shown in Fig. \ref{fig:springs} (a). Similar vibrational frequencies between the intra-molecular 
and inter-molecular modes allow for an effective exchange of energy between them, while a significant difference will tend to decouple 
the two sets of modes.
For this reason the rate of energy exchange between the translational mode ($T^{meso}$) and radial modes ($T^{rad}$) can be tuned by changing the inertia 
parameter($m^*$) which increases the overlap between the intra-molecular (sphere modes) and the translational phonon spectrum, and therefore, leads to a stronger coupling.
Figure \ref{fig:DoS} (a) shows the (un-normalized) vibrational spectrum for the breathing and the phonon modes for different values of the inertia parameter ($m^*$).  
Figure \ref{fig:DoS} (b) shows the mean squared frequency  $\langle \omega ^2 \rangle \equiv \int (2\pi \nu)^2 P(\nu) d\nu$, where $P(\nu)$ is the normalized density of states in Figure \ref{fig:DoS} (a). 
Note that the value of $m^*$ is chosen as to bring the mean of the sphere vibrational frequency to the same range as the mean of 
the phonon vibrational frequency. Table \ref{table:thetable} shows how these values depend on the compliance between the two springs.

\begin{table}\label{table:thetable}
\small
\caption{Parameters used in ChemDID}
\begin{tabular}{llcccccc}
\hline
& \multicolumn{2}{c}{\bf Internal Parameters} &  & & &\\
\hline
&parameter  & symbol & unit & \multicolumn{3}{r}{value}         \\
\hline
\hline
&Internal Heat  & \\
& Capacitance  &$C_{int}$  &---       & &  &  \multicolumn{1}{r}{60}         \\
& internal-to-intermolecular& \\
& coupling& $\nu_{meso}$ & 1/ps & & & \multicolumn{1}{r}{0.1}         \\
& internal-to-intramolecular& \\
& coupling& $\nu_{intra}$ & 1/ps& & & \multicolumn{1}{r}{0.1}         \\
& mesoparticle  mass & $m$ & g/mol& & & \multicolumn{1}{r}{296.16}         \\
\\
\hline
& \multicolumn{2}{c}{\bf Intermolecular parameters} &  & & &\\
\hline
%& \multicolumn{2}{c}{morse potential:} & & \multicolumn{4}{c}{$\phi_{inter}(R_{ij}-\sigma_i-\sigma_j;\epsilon_0,\gamma,r)$}  \\
&parameter  & symbol & units & \multicolumn{3}{r}{values}         \\
\hline
\hline
&morse potential  &\footnotesize{$\phi_{inter}(R_{ij}-\sigma_i-\sigma_j;\epsilon_0,\gamma,r_0)$}    & kcal/mol   & &  &  \multicolumn{1}{r}{Eqn. \ref{eqn:morse}} \\
&range  &$r_o$     & \AA   & &  &  \multicolumn{1}{r}{2.14}         \\
&curvature &$\gamma$ & ---  & & & \multicolumn{1}{r}{4.5}         \\
&energy &$\epsilon_o$& kcal/mol  &&& \multicolumn{1}{r}{7.0}         \\
& mean freq. square & $\langle \omega^2_{inter} \rangle$ &  1/ps$^2$&&& \multicolumn{1}{r}{120.0} \\
\\
\hline
& \multicolumn{2}{c}{\bf Intramolecular parameters\footnote{Parameters here have been defined such that $\Delta G > \Delta H$ and  $\Delta G > 5$ [kcal/mol]}} &  & & &\\
\hline
%&\multicolumn{2}{c}{ChemDID potential:} & & \multicolumn{4}{c}{$\phi_{intra}(\sigma;\sigma_{min},\sigma_{max},\Delta G(KK,A,\sigma_o),\Delta H)$}  \\
& parameter & symbol & unit & \multicolumn{4}{c}{values} \\
\hline
\hline
&ChemDID potential  &\footnotesize{$\phi_{intra}(\sigma;\sigma_{min},\Delta G,\Delta H)$} & kcal/mol   & &  &  \multicolumn{1}{r}{Eqn. \ref{eqn:intrapot}} \\
& barrier  &$\Delta G$ & kcal/mol & & 30 & - &  80          \\
& endothermicity  &$\Delta H$ & kcal/mol & & 0 & - &  20        \\
\\
\cline{2-8} 
& \multicolumn{3}{|c|}{parameters }  & \multicolumn{4}{|c|}{$\sigma_{min}$}  \\
\cline{5-8} 
& \multicolumn{3}{|c|}{depending on $\sigma_{min}$} & \multicolumn{1}{|c|}{1.50 \AA} & \multicolumn{1}{|c|}{1.75 \AA} & \multicolumn{1}{|c|}{2.00 \AA} & \multicolumn{1}{|c|}{ 2.25 \AA} \\
\cline{2-8}
\cline{2-8}
&\multicolumn{1}{|c}{volume change} & $(\Delta V/V_o)_{vdW}$ & --- &\multicolumn{1}{|c|}{65 \%} & \multicolumn{1}{|c|}{50 \%} & \multicolumn{1}{|c|}{35 \%} & \multicolumn{1}{|c|}{20 \%} \\
&\multicolumn{1}{|c}{inertial parameter} & $m^*$ & g/mol &\multicolumn{1}{|c|}{2961} & \multicolumn{1}{|c|}{4145} & \multicolumn{1}{|c|}{7402} & \multicolumn{1}{|c|}{21319} \\
&\multicolumn{1}{|c}{ mean freq.} & $\langle \omega^2_{rad} \rangle$ & 1/ps$^2$ &\multicolumn{1}{|c|}{118.4} & \multicolumn{1}{|c|}{123.00} & \multicolumn{1}{|c|}{120.3} & \multicolumn{1}{|c|}{119.2} \\
&\multicolumn{1}{|c}{curvature} & $K$ & --- & \multicolumn{1}{|c|}{80} & \multicolumn{1}{|c|}{253} & \multicolumn{1}{|c|}{1280} & \multicolumn{1}{|c|}{20500} \\
&\multicolumn{1}{|c}{Gaussian width} & $\sigma_o$ & \AA & \multicolumn{1}{|c|}{0.177} & \multicolumn{1}{|c|}{0.133} & \multicolumn{1}{|c|}{0.088} & \multicolumn{1}{|c|}{.044} \\
%cline{5-8}
%&\multicolumn{1}{|c}{Gaussian Const} & $A$ & kcal/mol &\multicolumn{4}{|c|}{$\Delta G - \frac{\Delta H}{2} - 5.0 $} \\
\cline{2-8}
\\
& Gaussian constant & $A$ &  kcal/mol&\multicolumn{4}{c}{$\Delta G - \frac{\Delta H}{2} - 5.0 $} \\
& maximum radius  &$\sigma_{max} $ & \AA &  & &   2.5         \\

\\
\end{tabular}
\label{table:thetable}
\end{table}

\section{Shock propagation and attenuation}

\subsection{Response to a sustained shock}

In the following we compare the shock response of inert and SWED materials using ChemDID.
The initial condition for the simulations consist of a target made of an fcc crystal obtained by 
replicating the four-atom fcc unit cell (with lattice parameter of 10.1 \AA) 200 times along the shock direction 
(z) and 20 times along the x and y directions leading to a system with 320,000 molecules.
Periodic boundary conditions are imposed along the x and y directions, while the z direction is open. 
The system is thermalized at 300 K for 100 ps bringing internal DoFs in ChemDID to thermal  equilibrium 
with radial and center of mass modes. 

\begin{figure}[h!]
  \centering
  %trims (crops) from left, bottom, right and top respectively
  \includegraphics[trim=0cm 2cm 0cm 1.3cm,clip=true,scale=1.15]{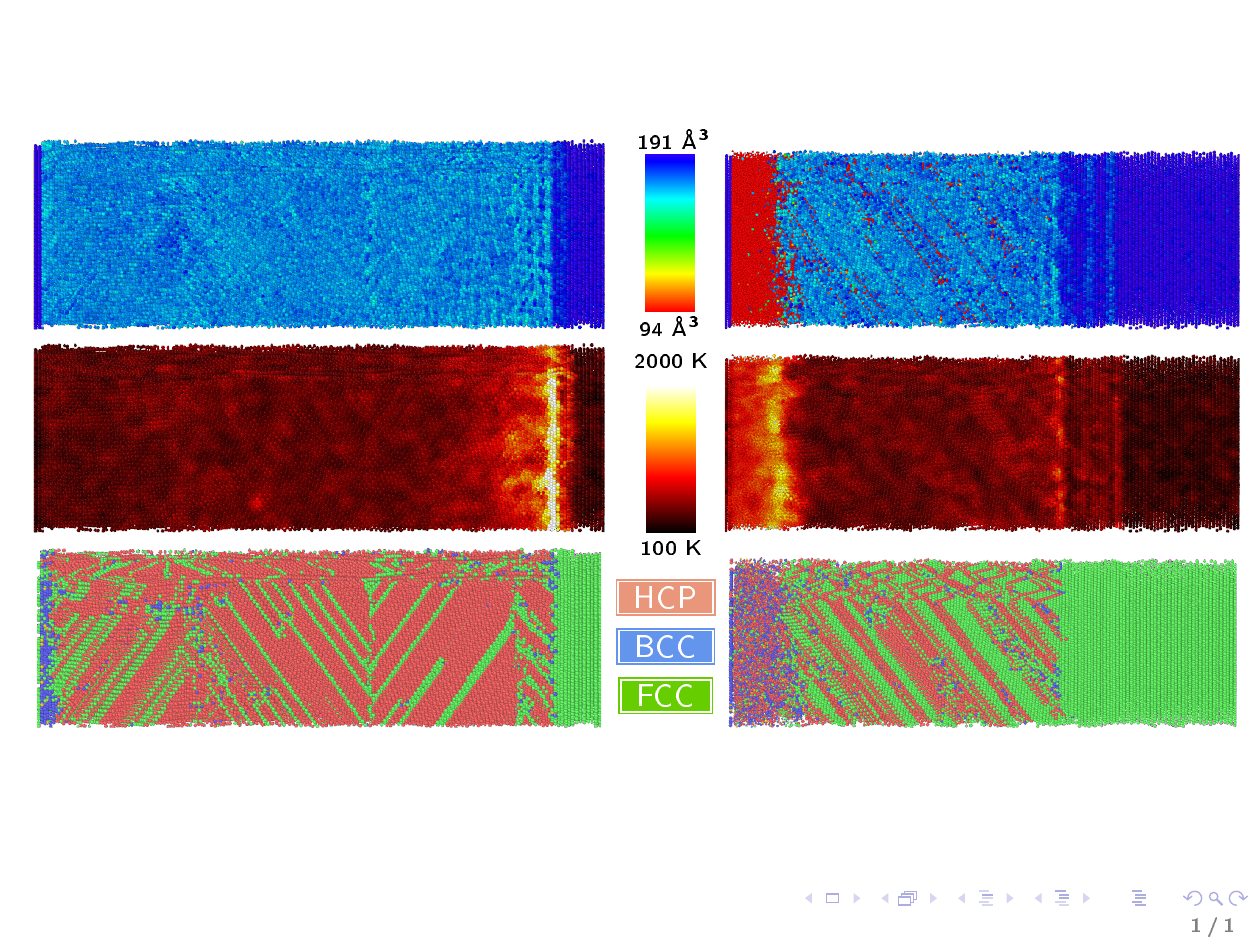}
  \caption{Snapshots showing (vdW) volume, temperature, and crystal structure for INERT (left) and SWED (right) materials impacted from the left by an infinitely massive piston with velocity of 2 km/sec}
  \label{fig:snapshots}
\end{figure}

The samples are impacted with a thin (one lattice constant thick), rigid and infinitely massive piston traveling at 
the {\it piston velocity} (u$_p$). For piston's speeds below 2.0 km/sec we use an integration time step of $dt = 0.005$ ps, whereas
for speeds above this value we halve the time step in order to get numerical stability. 
The infinitely massive piston does not slow down due to the interactions with the 
target, nor rarefaction waves are generated from its free surface. Thus this setup generates a {\it sustained shock} 
as opposed to one of finite duration obtained with finite pistons.
Since the piston does not slow down, the material that is in its immediate proximity will quickly approach the same velocity u$_p$. 
The material under stress will attempt to deform or rearrange in order to alleviate its local stress build up. 
While in an inert material, plastic deformation or phase transformations are the only mechanisms that allow for stress relaxation, 
in a SWED material volume-reducing chemical reactions can significantly reduce pressure build up and attenuate the leading 
shock, as we shall see. 

Figure \ref{fig:snapshots} shows atomic snapshots of steady shocks on a SWED and inert materials shocked with u$_p$=2.0 km/s. Molecules 
are colored to indicate local volume, temperature, and crystal structure. Interestingly, the SWED material exhibits a region of reacted 
(volume collapsed) material next to the piston; this reacted region propagates along the shock direction. As expected, in the inert case 
shocked particles also reduce their volume due to the high pressures but this occurs rather uniformly throughout the 
sample. Importantly, we see that the leading shock has propagated much further in the inert material (all snapshots correspond to
the same time of 25 ps), an indication that the chemistry is weakening the shock.
The amount of plastic deformation following the shock also sheds light into its strength. We analyze plastic and structural transformations 
by characterizing the local crystal environment of each molecule. Hcp atoms in the fcc crystal indicate stacking faults that separate partial 
dislocations; thus red atoms indicate the traces of partial dislocations or regions that transformed to hcp. We see significantly less plastic deformation 
in the SWED material; this is because the leading shock is weaker. A temperature spike follows the primary wave in the 
inert case, but it in the SWED case the highest temperatures localize in the chemically reacted region. 

\begin{figure}[h!]
  %\par\vspace{-10mm}
  \centering
  \begin{tabular}{cc}
    \hspace{-1.0cm}\includegraphics[scale=0.5]{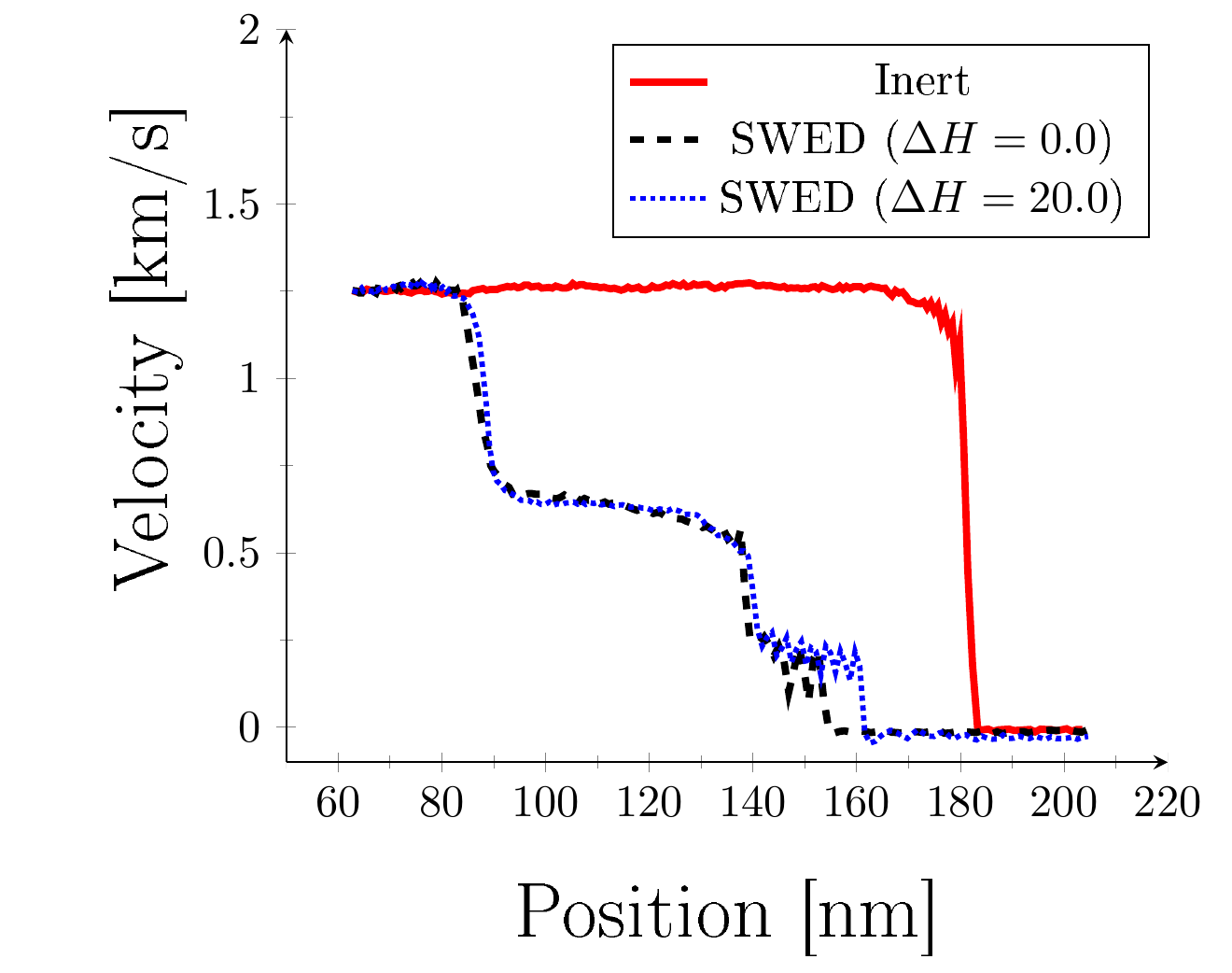} &
  \includegraphics[scale=0.5]{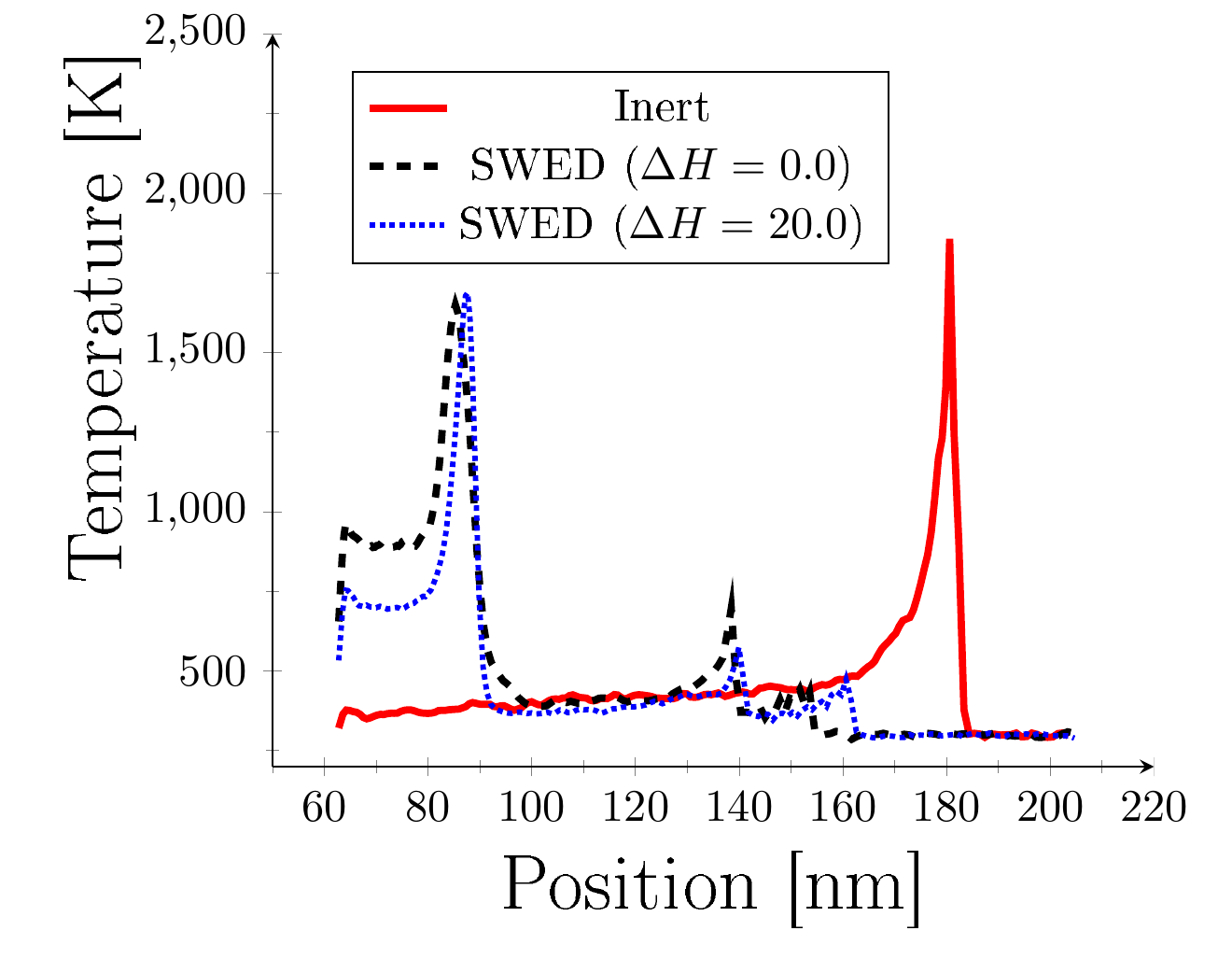} \\
  \includegraphics[scale=0.5]{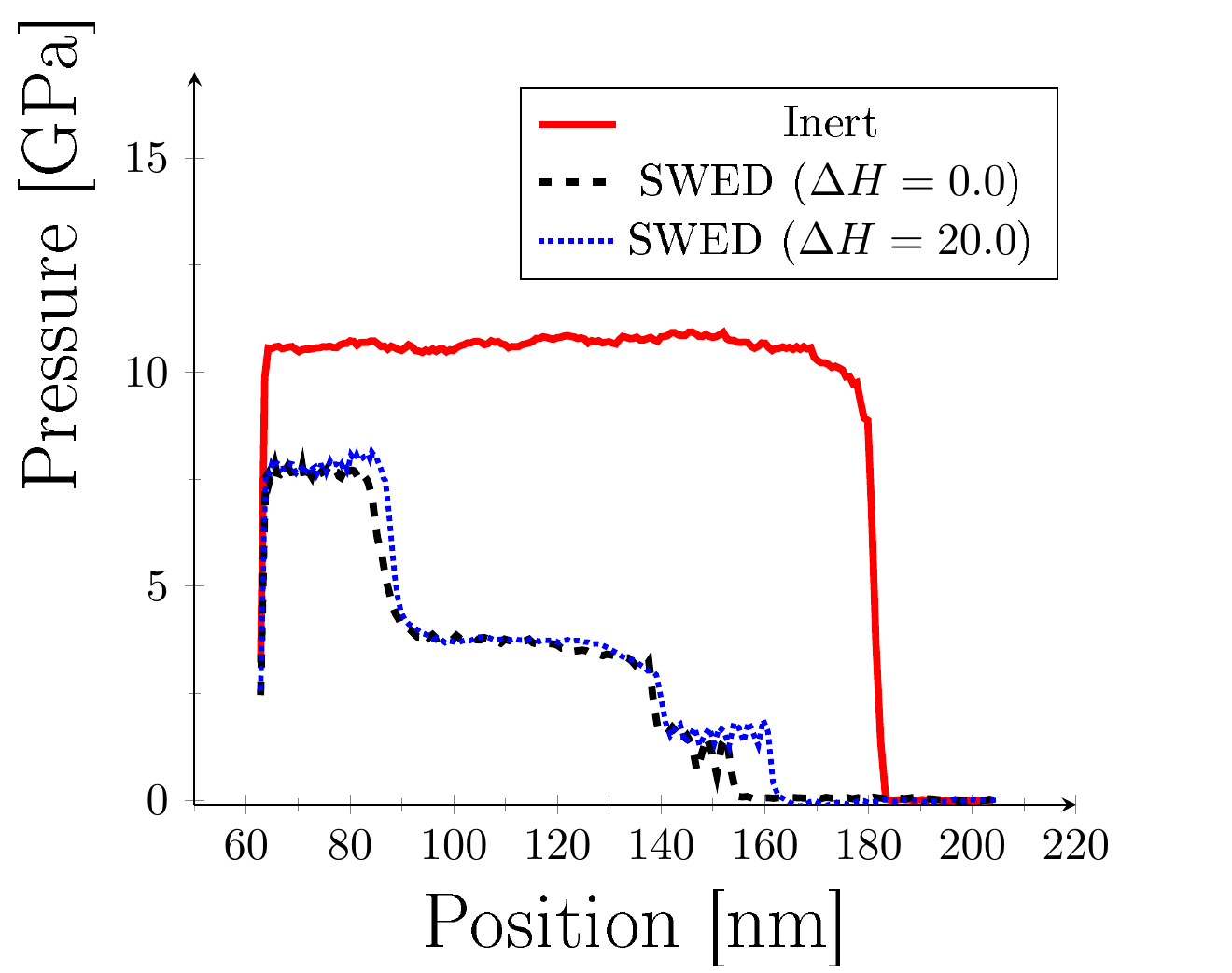} &
  \includegraphics[scale=0.5]{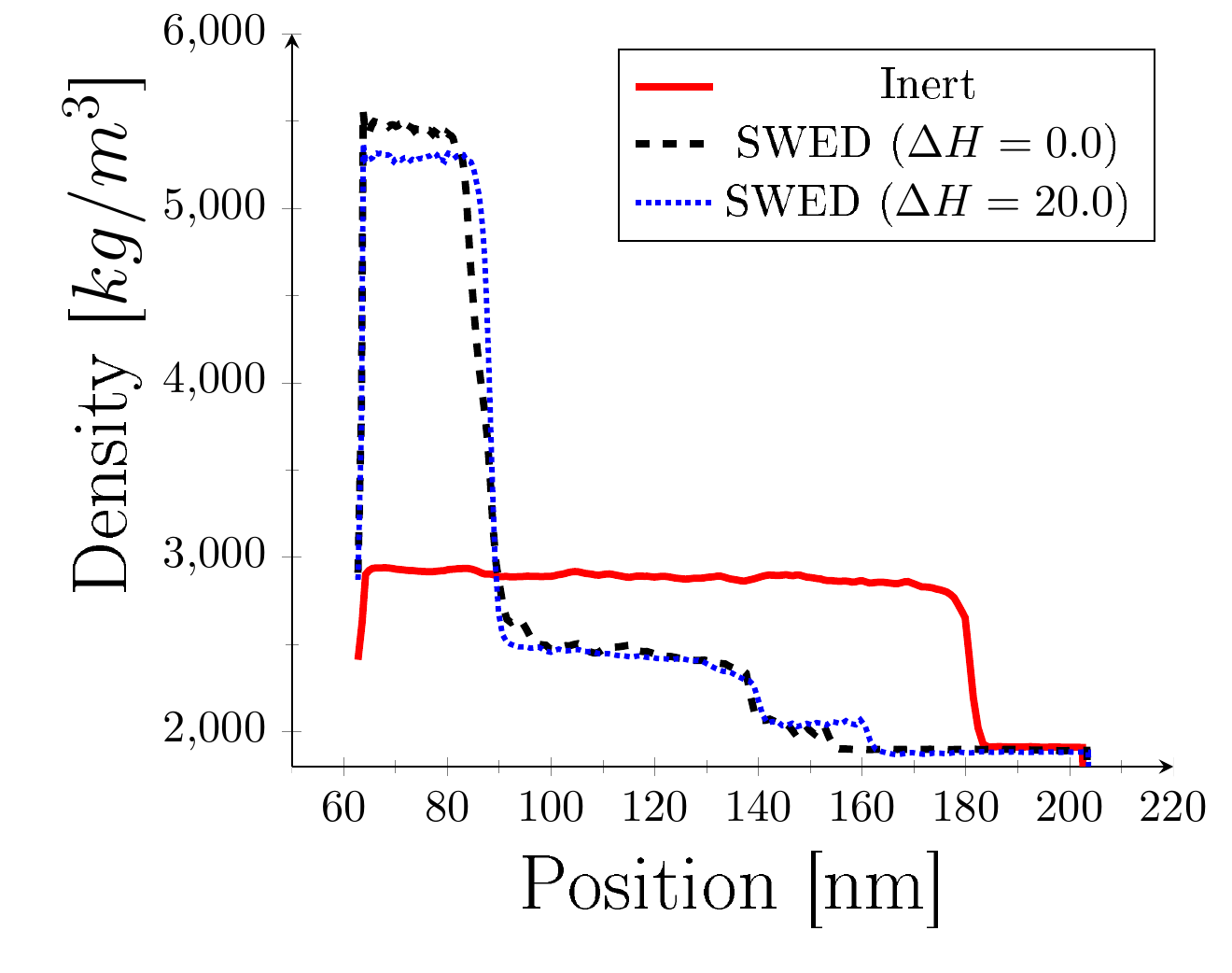} \\
  \end{tabular}
  \caption{ Velocity, Temperature, Pressure , and density profiles for a inert (solid) and swed (dashed/dotted) samples. 
    The piston speed is $u_p$ = 1.25 [km/sec]}
  \label{fig:Profiles}
\end{figure}

\begin{figure}[h!]
  %\par\vspace{-10mm}
  \centering
  \begin{tabular}{cc}
  \includegraphics[scale=0.534]{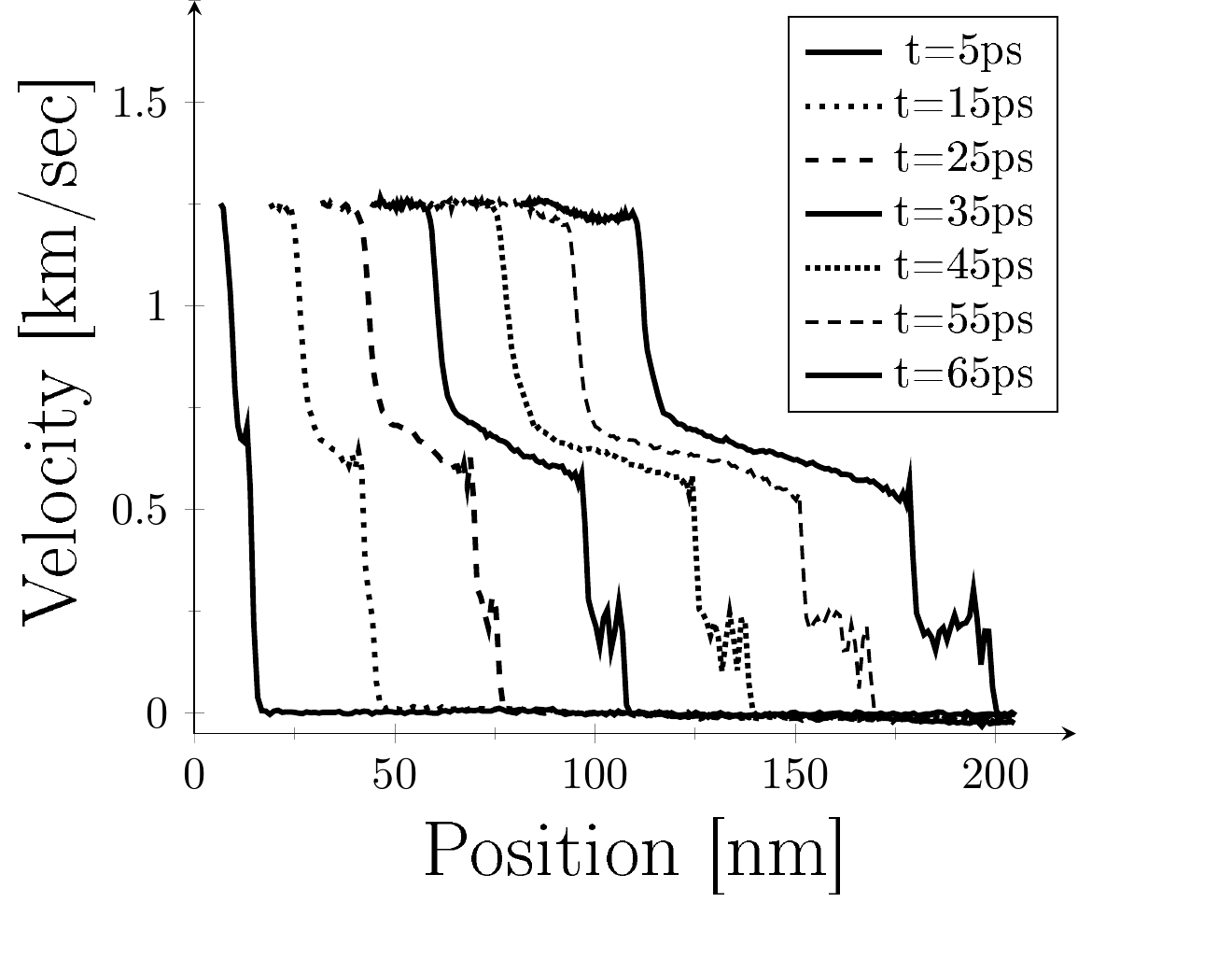} &
  \includegraphics[scale=0.534]{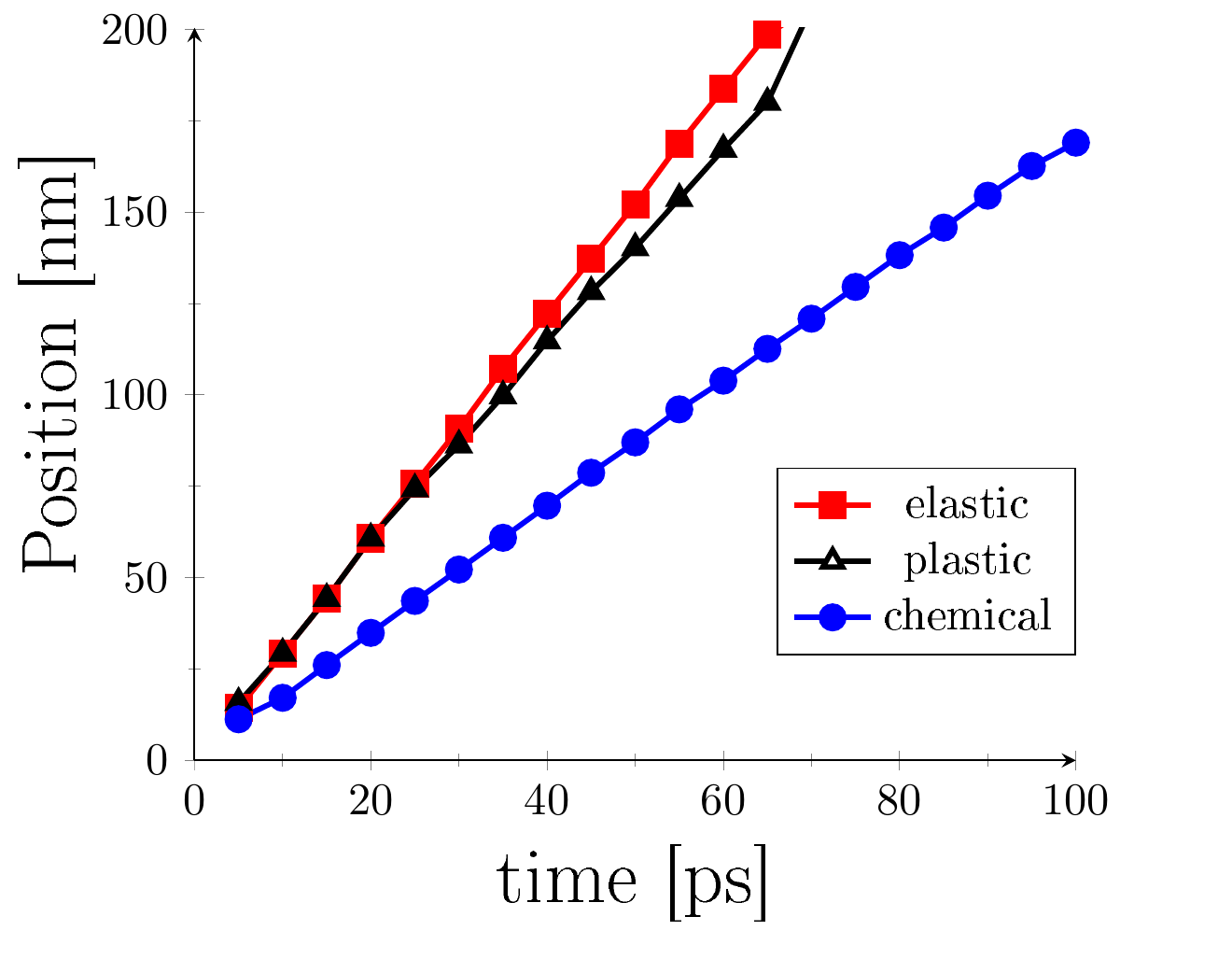} \\
  \end{tabular}
  \caption{ (LEFT) Velocity profiles at different times, (RIGHT) Position of the shock fronts vs time}
  \label{fig:wavefronts}
\end{figure}

Figure \ref{fig:Profiles} contrasts the profiles of key thermodynamic quantities for a SWED and inert materials at a given time; 
the SWED profiles further contrast the response with zero ($\Delta H=0$ kcal/mol) and finite endothermicity ($\Delta H = 20$ kcal/mol).
A three-wave structure can be seen for the SWED material. The leading wave (often called elastic precursor) shocks the
material and behind it a plastic wave propagates at a slower velocity. Following the plastic is a chemical wave that
densifies and heats up the material. Note that higher endothermicity decreases the temperature in the reacted region. Also 
the stiffness here will be slightly higher due to the linear term in $\Delta H$ appearing in the intramolecular potential.
It is interesting to note that the pressure is greater in the reacted material than in the rest of the sample. 
It will be shown below, that this is a consequence of the volume collapse during a sustained shock and a necessity for steady state shocks.
A similar three-wave structure has been observed experimentally on porous copper \cite{Boade}.

Figure \ref{fig:wavefronts} (a) shows velocity profiles at different times and depicts the
development of the three wave structure. The chemical wave separates from the initial shock at short times and the plastic and elastic 
waves take significantly longer time to separate. For the case shown the separation between 
the elastic and plastic waves occurs at around 30 ps, after which the two waves trail each other within a few nm. 
In contrast, the inert material is in an overdriven regime where the plasticity has merged with its elastic propagation into a single wave. 
Figure \ref{fig:wavefronts} (b) shows the position of the three wave fronts as a function of time from which we obtain their velocities.
We see that steady-state is reached very early for the chemical wave, whereas for the plastic/elastic fronts it takes a longer time for this to occur.

\subsection{Taming shockwaves with volume-reducing chemical reactions}

Now that we have established the possibility of weakening shocks with chemical reactions that involve volume collapse, 
we are interested in understanding how the characteristics of the chemistry affect the strength of the coupling and the
SWED effect. Before describing our ChemDID results we use conservation laws to develop a general framework to discuss SWED and 
understand how volume collapse, activation barrier and endothermicity contribute to achieving the desired effect.

\begin{figure}[h!]
  \centering
  \hspace{2.5cm}\includegraphics[trim=1cm 6cm 9cm 4cm,clip=true,scale=.75]{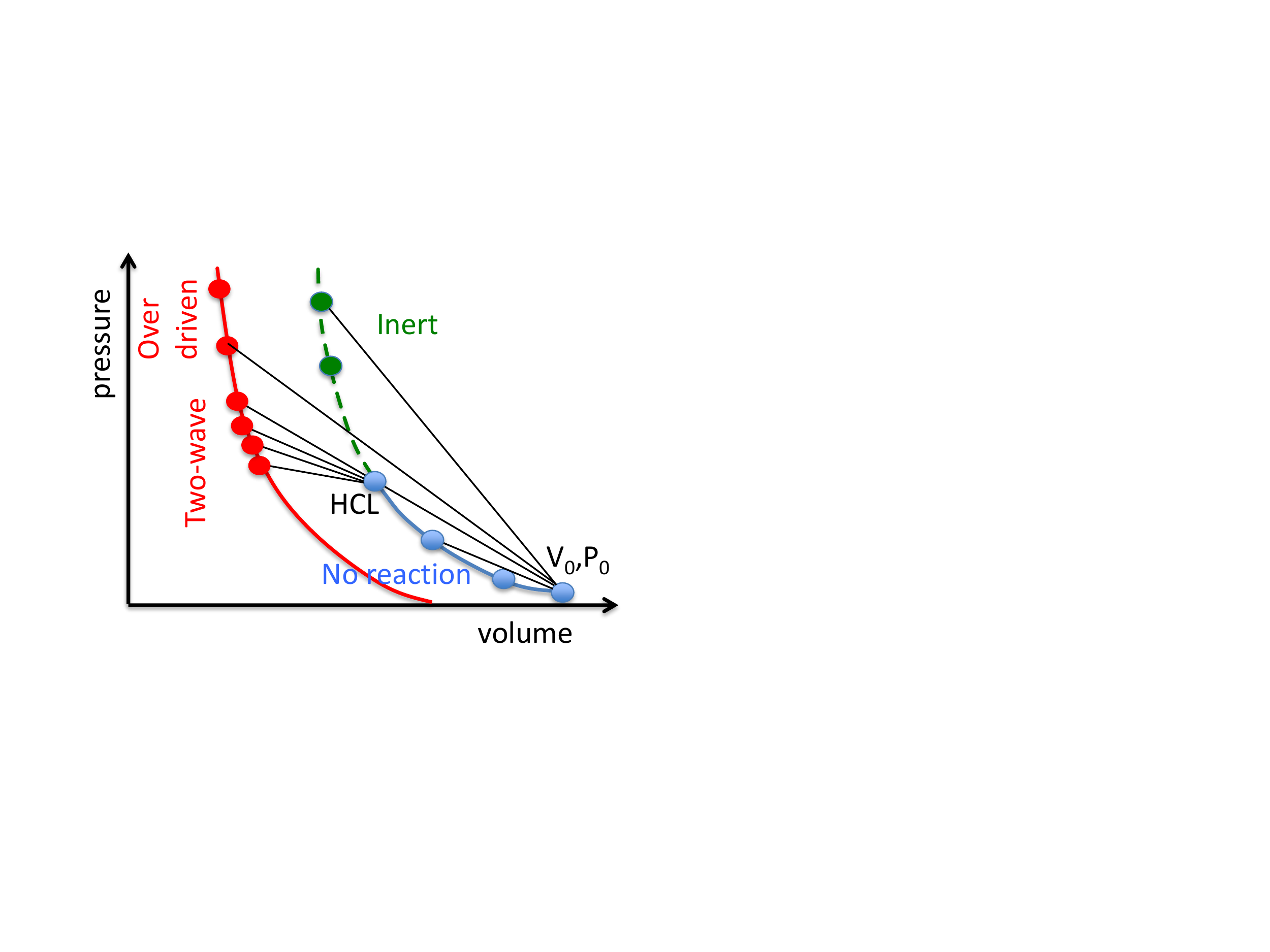}
  \caption{Schematic showing the paths in the P-V space for inert and reactive shocks.
    The critical Hugoniot Chemical Limit (HCL) connects the transition point between the two Hugoniot curves. 
  }
  \label{fig:sketch}
\end{figure}

We start by considering the locus of the pressure-volume states accessible by shocking a material initially
at volume $V_0$ and negligible pressure $P_0\sim0$, see blue dots in Figure \ref{fig:sketch}. It is important
to understand that these states (the Hugoniot of the material) is different from a reversible equation of state;
each point is the result of a shock experiment and can only be accessed by shocking material at the initial point  ($V_0$,$P_0$).
Assuming the shock is too weak to trigger chemistry or plastic deformations, we will have a single wave structure \cite{GermannPRL2000}.
By applying mass and momentum conservation across the shock front we will obtain interesting insight into the SWED behavior.

Requiring mass conservation across the shock wave \cite{ShockBook} yields the following equality:
\begin{align}
  \rho_s (\dot{x}_s - u_s)  & =   \rho_0 (\dot{x}_0 - u_s)  
\end{align}
where $\rho_s$ and $\dot{x}_s$ represent the density and  particle velocity in the shocked region and, as before, subscript 0 denote the 
unshocked region. Since the unshocked material is not moving ($\dot{x}_o=0$), we can readily solve for the particle velocity in the shocked 
region in terms of the densities and the shock velocity:
\begin{align}\label{eqn:vs}
  \dot{x}_s &= u_s ( 1- \frac{\rho_0}{\rho_s}).
\end{align}

An expression for the pressure in the shocked region can be written from the momentum conservation equation as:
\begin{align}\label{eqn:RH1}
P_s = \rho_o u_s \dot{x}_s = \rho_o u_s^2 (1 - \frac{\rho_o}{\rho_s})
\end{align}
The case when the pre-shocked region is not stationary is given in the appendix \ref{sec:apdxB}.
Rearranging the two conservation equations, known as Rankine-Hugoniot equations,\cite{ShockBook} we can write down an expression for the shock velocity
in terms of the change in pressure and volume 
\begin{align}\label{eqn:RH2}
 \rho_o^2 u_s^2 = - \frac{P_s - P_0 }{V_s - V_0} 
\end{align}

We see that the slope of the line connecting the initial and shocked materials (black lines in Fig. \ref{fig:sketch} known
as Rayleigh lines) is related to the velocity squared of the corresponding wave. It is clear from the conservation equations that in 
order to decrease the pressure following the initial shock it is necessary to slow down the shock speed $u_s$ or, equivalently, 
maintain the density in the shocked region as close as possible to the un-shocked density.
Let's delve into how chemical reactions can help to maintain both of these conditions.

If we shock the material above a given threshold, that we will call Hugoniot chemical limit (HCL) in analogy to the Hugoniot elastic limit, 
a chemical wave will develop behind the elastic one. For simplicity we will consider two wave structures and neglect the plastic wave. 
In the pressure-volume plot we connect the state representing the elastic wave with the Hugoniot of the chemical
products (red circles in Fig. \ref{fig:sketch}). We can see that there will be a series of shocks for which 
the velocity of the chemical wave will be lower (gentler slopes) than that of the leading elastic precursor (steeper slope).
In this regime, the point ($P_{HCL},V_{HCL}$) in the Hugoniot determines the propagation velocity and pressure of the elastic precursor
through Eqn. (\ref{eqn:RH1}) and (\ref{eqn:RH2}), where s = HCL (assuming no plasticity).
This value is independent of the piston velocity (the dependence of the HCL with the activation barrier is described in the next section) until
the velocity of the chemical wave matches that of the elastic precursor. At this point, the overdriven regime is reached, and the profile of 
the wave is characterized by a single wave structure.

Let us now quantify how the chemical wave weakens the leading shock wave. Applying mass conservation across the chemical 
wave \{ c \}, i.e. between the reacted and shocked regions leads to:
\begin{align}\label{eqn:vc}
  \rho_c (\dot{x}_c - u_c)  & =   \rho_s (\dot{x}_s - u_c) 
\end{align}
where $\rho_c$ and $\dot{x_c}$ represent the density and particle velocity in the reacted region.
Substituting Eq. \ref{eqn:vs} into Eq. \ref{eqn:vc} and rearranging terms we  obtain an expression for the shock speed as:
\begin{align}\label{eqn:xsdot}
  u_s & = \frac{\dot{x}_s}{1 - \frac{\rho_o}{\rho_s}}  = \frac{\frac{\rho_c}{\rho_s}(\dot{x}_c - u_c) + u_c}{1 - \frac{\rho_o}{\rho_s}} 
\end{align}
Moreover, in a sustained shock the particle velocity of the reacted material is equal to the piston velocity ($\dot{x}_c = u_p$). Hence, we arrive at an 
expression for the pressure following the leading shock in terms of the densities of the various regions and the velocity of the chemical wave:
\begin{align}\label{eqn:Ps}
P_s = \rho_o u_s \dot{x}_s = \rho_o \frac{(\frac{\rho_c}{\rho_s}(u_p - u_c) + u_c)^2}{1 - \frac{\rho_o}{\rho_s}}
\end{align}

Let us comment on the regime under which this expression is valid. In order for waves to separate into a chemical and a shocked 
components, the velocity of the chemical wave needs to lie between the piston velocity $u_p$ and the wave speed corresponding to
the HCL: $u^{HCL}_s$.  
The upper limit, $u_c = u^{HCL}_s$, corresponds to the overdriven regime when chemical wave and the shock wave travel at a single 
speed $ u^{*}_s = \frac{u_p}{1 - \frac{\rho_o}{\rho_c}} $. The lower limit, i.e. a chemical wave traveling at $u_c \leq u_p$  corresponds to
no chemistry production, viz. $u_c \rightarrow 0 \Rightarrow \rho_c \rightarrow \rho_s$, therefore the (inert) shocked wave travels at speed: 
$u_s^I = \frac{u_p}{1 - \frac{\rho_o}{\rho_s}} $.

\begin{figure}[h!]
  \centering
  %trims (crops) from left, bottom, right and top respectively
  \begin{tabular}{cc}
  \includegraphics[trim=0cm 0cm 0cm 0cm,clip=true,scale=.62]{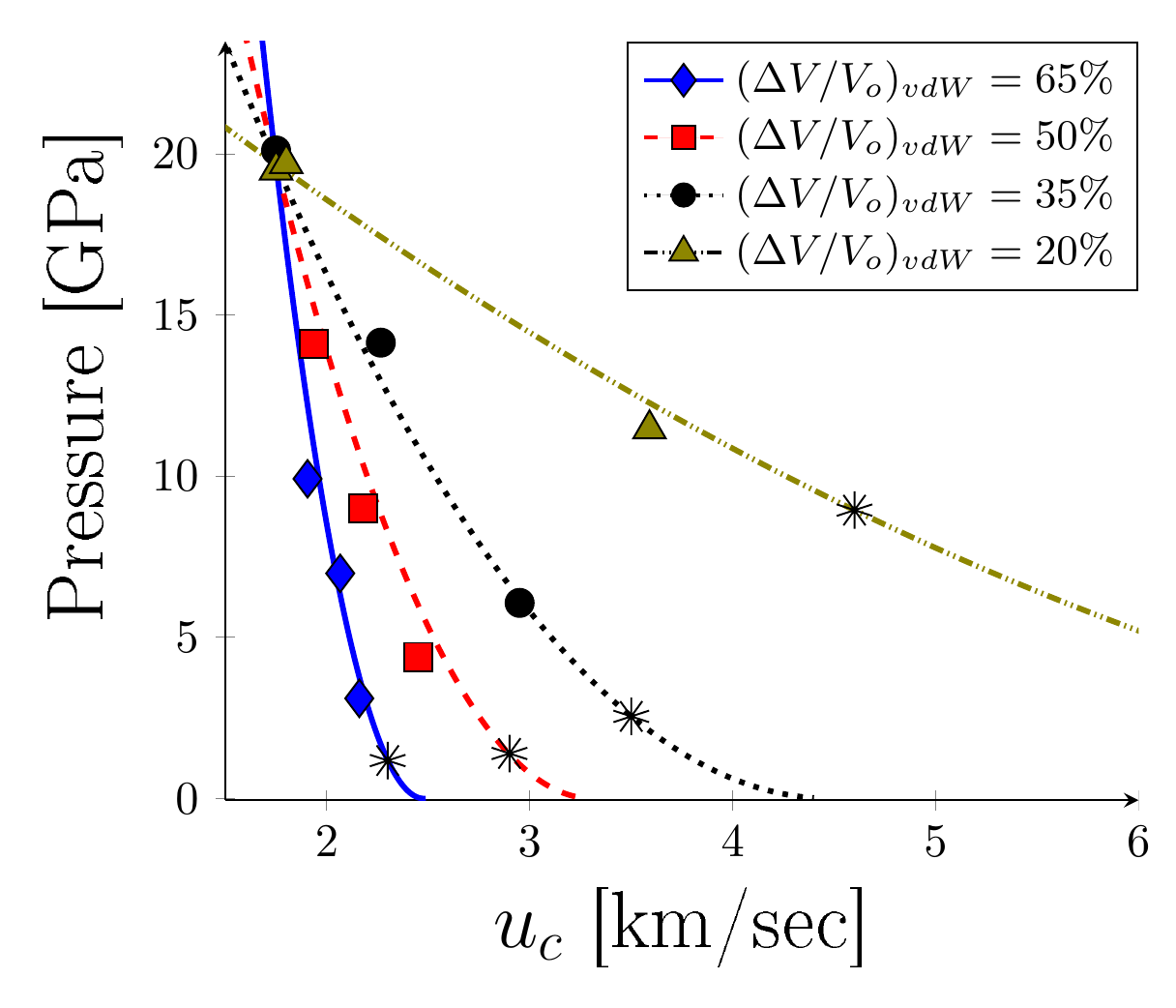}&
  \includegraphics[trim=0cm 0cm 0cm 0cm,clip=true,scale=.6]{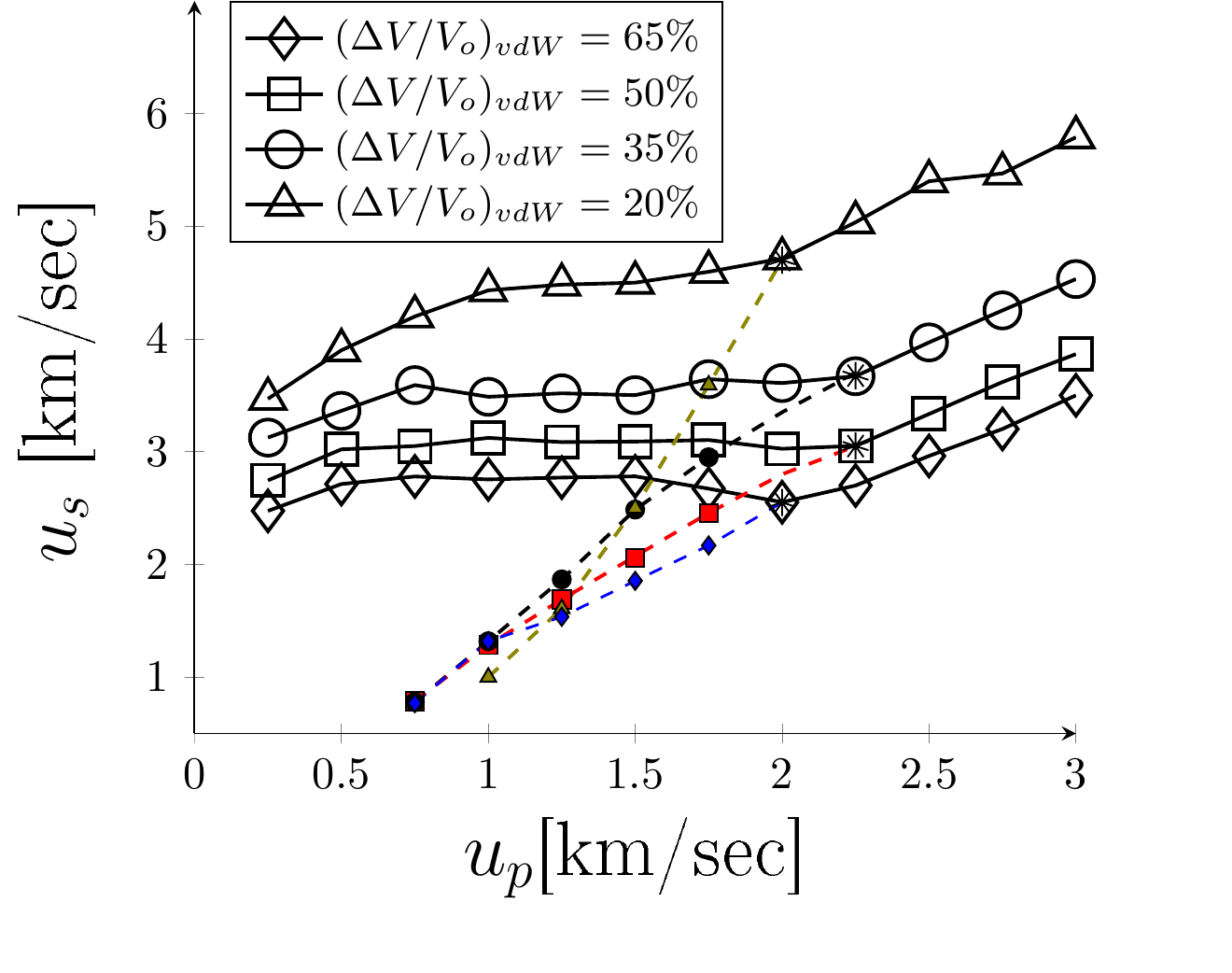} \\
  (a) & (b) \\
  \end{tabular}
  \caption{(a) Expression \ref{eqn:Ps} (for $u_p$=1.75 km/sec) compared to simulated data with different volume and chemical speeds;
    for a given volume collapse, a lower activation barrier corresponds to a faster chemical speed. The point where the chemical wave reaches the overdriven regime is denoted by ($\ast$). (b) Shock speed vs impact speed for various volumes (for the same activation barrier $\Delta G = 30$ kcal/mol); The chemical waves are denoted by the dashed lines.
}
\label{fig:Ps}
\end{figure}

Equation \ref{eqn:Ps} shows that in order to weaken the shock, i.e. reduce $P_s$, we need $\rho_c$ to be high 
(large volume collapse) but also we need the chemical wave to travel at fast speeds. This makes it clear that, as discussed earlier, reaction kinetics
are critically important for SWED. We now compare the predictions of Eq. \ref{eqn:Ps} for the pressure behind
the leading shockwave with results from explicit ChemDID for a family of SWED materials with various volume collapse
amounts and activation energies. The density in the shocked regions $\rho_s$ depends slightly on the activation barrier, 
but we will assume it to be constant at $\rho_s = 2500$ kg/m$^3$ to evaluate Eq. \ref{eqn:Ps}; this number changes at most by 20 \% for all the cases 
considered at this piston speed. We shall see that this assumption leads to reasonable predictions and, in the next section, we will explore the 
sensitivity of the results with respect to it.

Figure \ref{fig:Ps} (a)  shows the pressure following the initial shock as a function of the velocity of the chemical wave. The lines show the 
predictions from Eq. \ref{eqn:Ps} and the symbols represent ChemDID results. The results are shown for four model materials with different 
amounts of volume collapse and various activation barriers ($\Delta G =  \{80,60,30\}$ kcal/mol). As expected, increasing the amount of volume 
collapse results in lower shock pressures. The non-linear dependence of pressure on the density of the compressed state is consistent with the 
analysis of shocks on porous materials \cite{Gogulya1985shock,davison1971shock}. 
For a given volume collapse, lowering the activation energy (that controls reaction kinetics) leads
to faster chemical waves and also result in a reduction of the shock pressure. Figure \ref{fig:Ps} (b) shows the effect of volume-collapse on the 
leading shockwave (open symbols) and chemical wave (filled symbols) speed as a function of the piston velocity; we show the same volume-collapse 
cases in (a). Higher volume collapse result in lower shock velocities for all piston velocity and a reduction of the velocity corresponding to the HCL.
Once a chemical wave propagates, it couples with the leading shock wave resulting in a plateau of the shock wave speed as a 
function of the piston speed. Interestingly, we find that chemical waves tend to propagate faster the smaller the volume-collapse for a given
piston velocity. These results are also consistent with the shock experiments on porous copper of different densities\cite{Boade}.
The authors report a three-wave structure, similar to what is observed in our simulations, where measurement of the slower wave (equivalent to our 
chemical wave) shows a steeper slope for samples with less porosity. The point where the chemical wave merges with the leading shock wave $u^{HCL}_s$ 
can also be seen in Figures \ref{fig:Ps} (a) and (b) and is denoted by the asterisks. This points marks the end of the shock absorbing capabilities of the
material and further increasing the piston velocity leads to increased shock velocity and pressure. 

In summary, we have shown that volume collapse is the dominant factor in determining the effectiveness of the chemical reactions at weakening the 
leading shockwave (i.e. reducing the pressure) but reaction kinetics is also important. On the other hand, large volume collapses cases exhibit HCLs 
corresponding to slower shock velocities and the design of materials for shock absorption should take into account the regime of operation.

\subsection{Quantifying the role of chemistry on shock propagation}

As discussed in the previous sub-section, in order to engineer materials with desired shock-wave dissipation response, it is imperative 
to establish relationships between the characteristics of the chemical reactions and the material response.
We have already shown the importance of volume collapse and the velocity of the chemical wave. 
In this section we explore ChemDID simulations in more detail, in order to correlate microscopic characteristics of the chemical reactions
with shock dissipation.

\begin{figure}[h!]
  \centering
  \begin{tabular}{cc}
    \hspace{-1.5cm}\includegraphics[scale=.675]{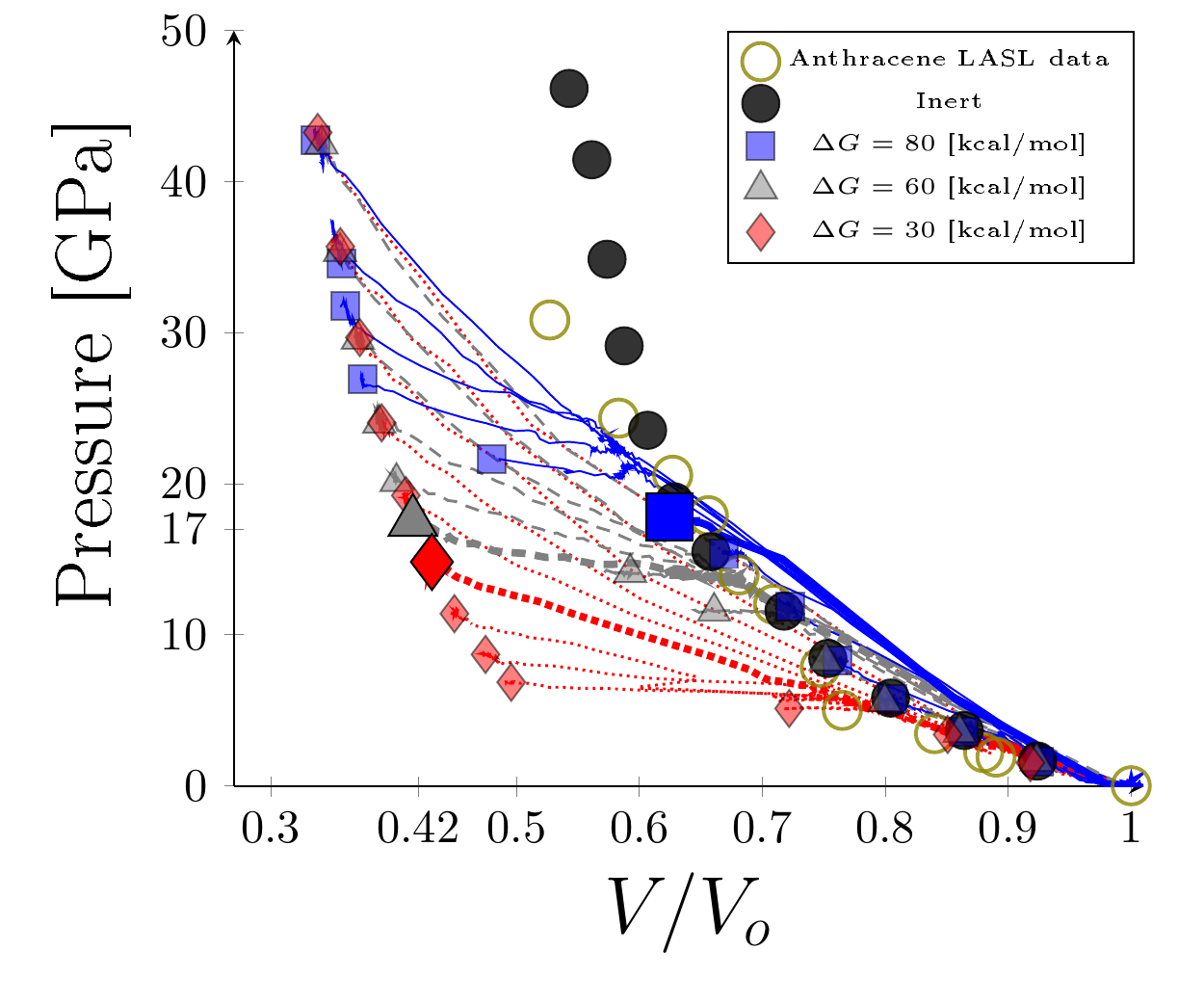} &
    \includegraphics[scale=.675]{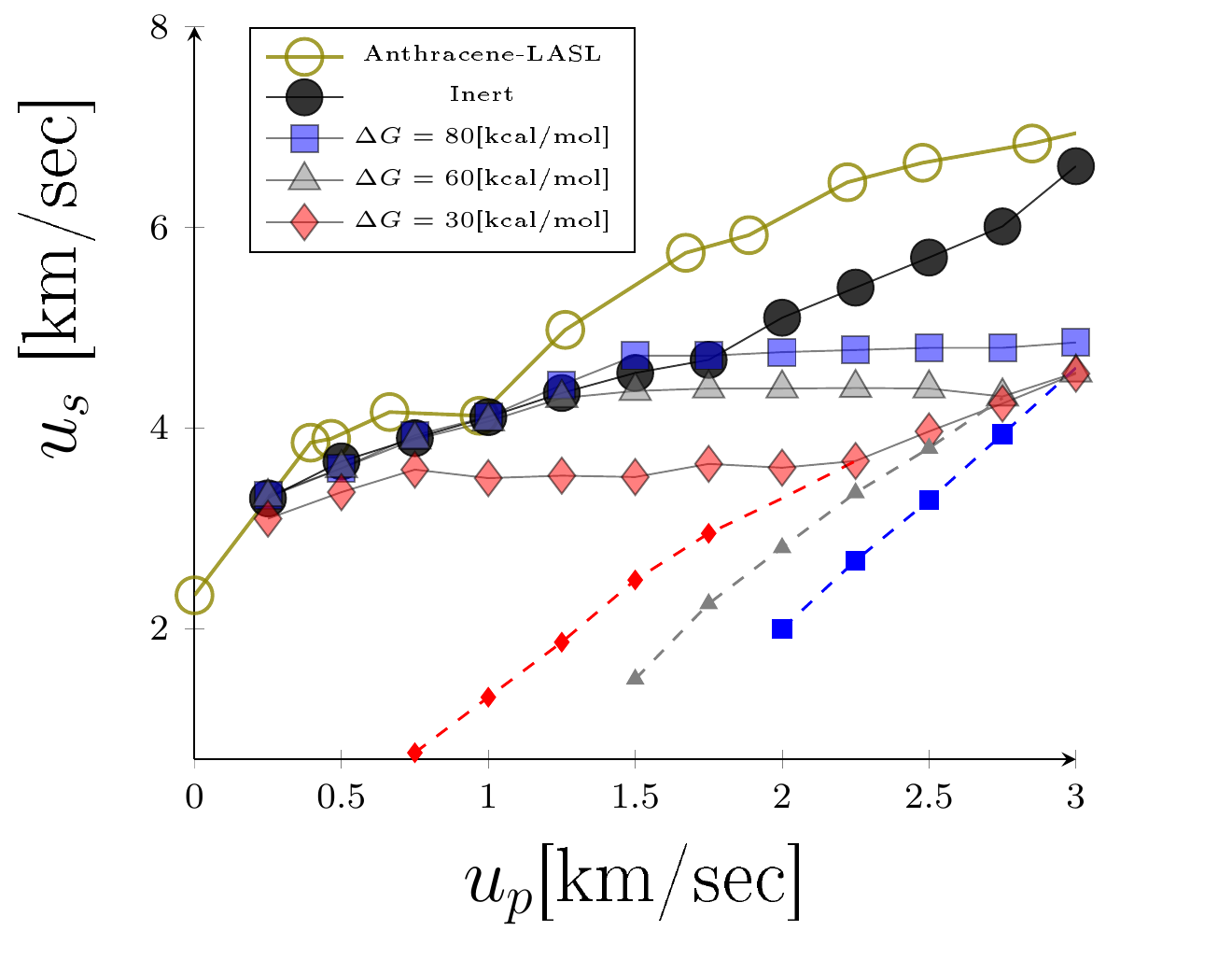}  \\
    (a) & (b) 
  \end{tabular}
  \caption{(a) Hugoniot and (b) Us-Up curves for an Inert and SWED materials with a curvature on the intramolecular potential corresponding 
  to $(\Delta V/V_0)_{vdW} = 35$ \%.  The different symbols correspond to different barriers. A single impact speed $u_p$=  1.75 km/sec has 
  been emphasized in (a) with larger symbols and thicker lines. The chemical waves are shown in dashed lines in (b).
   }
    \label{fig:barrierfigs}
\end{figure}

To understand the development of a two-wave structure and its properties we show the evolution of the system
in the pressure-volume space. Using ChemDID, we follow the local properties of a thin slab of material during shock loading.
The results are shown in Fig. \ref{fig:barrierfigs} (a) for three different activation barriers for a volume collapse of $(\Delta V/V_0)_{vdW} = 35$ \%.
This figure shows the individual paths for a range of piston speeds between $ u_p = {0.25 - 3.0}$ km/sec (in steps of 0.25 km/sec). 
For velocities below a critical value (HCL-limit; see Fig \ref{fig:sketch}), the ends of the Rayleigh lines follow the inert (unreacted) Hugoniot. 
After a critical compression is reached, corresponding the volume of the HCL $V^{HCL}_s$ ($ =1/\rho^{HCL}_s$) which depends slightly on the 
activation barrier for chemistry, the system develops a two-wave structure described by two Rayleigh lines: the first going from the
initial state to the pressure and volume after the initial wave (along the unreacted Hugoniot) and the second crossing into the reacted Hugoniot.
In this description we are combing the plastic and elastic waves into one as their propagation velocities are similar and they do not
separate significantly during the simulation time.

In order to investigate the dominant factors during this transition, we can write down the following expression 
relating the pressures in the reacted region $P_c$ and unreacted shocked region $P^{HCL}_s$; the derivation is shown
in Section \ref{sec:apdxB} of the Appendix.
\begin{align}\label{eqn:Pc}
  P_c - P^{HCL}_s  = &  (u_c - \dot{x}_c)^2 \rho_c (\frac{\rho_c}{\rho^{HCL}_s} - 1 )
\end{align}
The pressure in the reacted zone $P_c$ will lie in the reacted Hugoniot and will depend primarily on the impact speed.
Higher pistons will cause the reacted region to move towards higher pressures along the product Hugoniot.
This can be seen in Fig.\ref{fig:barrierfigs} where the case $u_p$=1.75 km/sec has been emphasized. The pressure response 
of the reacted material for this case will tend to collapse to the same region around $V/V_0 = .42$ and P = 17 GPa. 
The critical pressure $P^{HCL}_s$, on the other hand, will depend on the onset of chemistry. As seen from  Eqn. \ref{eqn:Pc}, this pressure 
is mainly dominated by density of the reacted material and the propagation velocity of the chemical wave since they appear quadratic in this 
expression, while the critical volume $1/\rho^{HCL}_s$ appears linearly.  
This is the reason we were able to ignore the crucial density ($\rho^{HCL}_s$) in Eqn. \ref{eqn:Ps} and obtain and expression 
that matches well with the simulated data. 
We also see from Eqn. \ref{eqn:Pc},  that the change in pressure cannot be negative, since $\rho_c>\rho^{HCL}_s$. The slope of the Raleigh line 
in the P-V plane is therefore zero near threshold, and negative for case with more volume collapse. The slope of the Raleigh-line determines 
the propagation velocity of the chemical front. The points that lie between the reacted and inert Hugoniots represents cases with heterogeneous 
regions of collapsed volume which do fully propagate, at least within the scales of our simulation.

In Fig. \ref{fig:barrierfigs} (b) we compare the shock velocities for an inert an reactive ChemDID simulations with various activation energies and 
includes experimental data on anthracene for reference \cite{LASL}. 
In the elastic regime, i.e. shocks weaker than the HEL, the shock speed increases linearly with  piston velocity. 
The HEL is reached for piston velocities of approximately 0.75 km/sec, this value depends weakly on the stiffness of the intra-molecular potential;
stiffer materials tend to have slightly higher HEL limits. The HCL, on the other hand, occurs at slightly higher piston velocities and 
depends strongly on the activation barrier of the intra-molecular chemistry (dashed lines) in Fig. \ref{fig:barrierfigs} (b).
The case where HEL $\approx$ HCL (occurring for small barriers) is studied in more detail in section \ref{sec:last}, where we discuss how 
chemistry directly affect the different energy transferring mechanisms.
We find that lowering the activation barrier on the chemical reactions leads to faster chemistry and a stronger coupling to the leading shock wave, 
and therefore, to a reduction in the pressure in this region. Whereas the propagation of the chemical and elastic waves achieve steady-state during 
the simulation, an accurate determination of the velocity plastic wave is difficult for the sizes of the system simulated as the plastic wave separates 
from the elastic precursor at late times, see Fig. \ref{fig:wavefronts} (b).

\begin{figure}[h!]
  \centering
  \begin{tabular}{cc}
    \hspace{-1.2cm}\includegraphics[scale=.675]{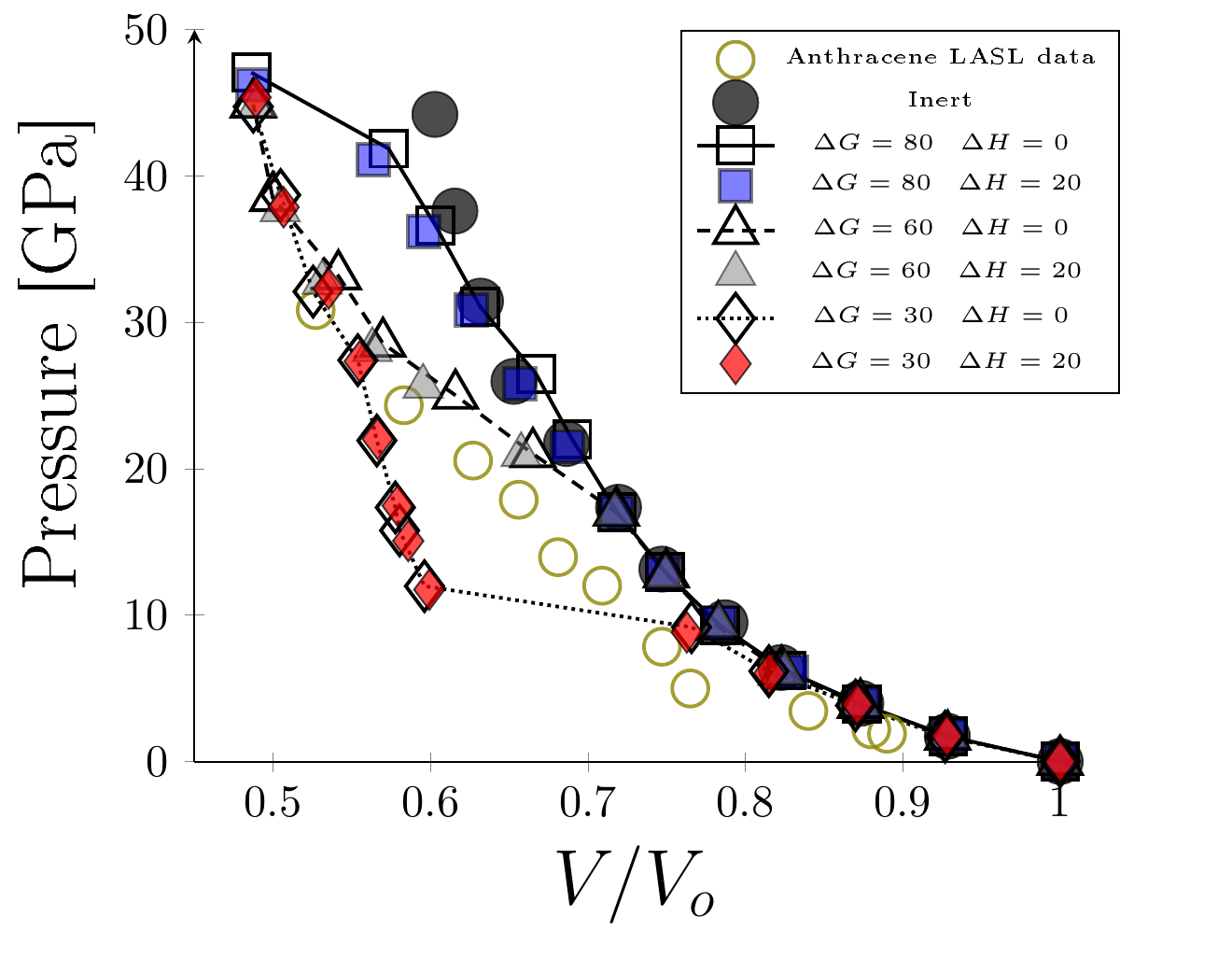} &
    \hspace{-.75cm}\includegraphics[scale=.675]{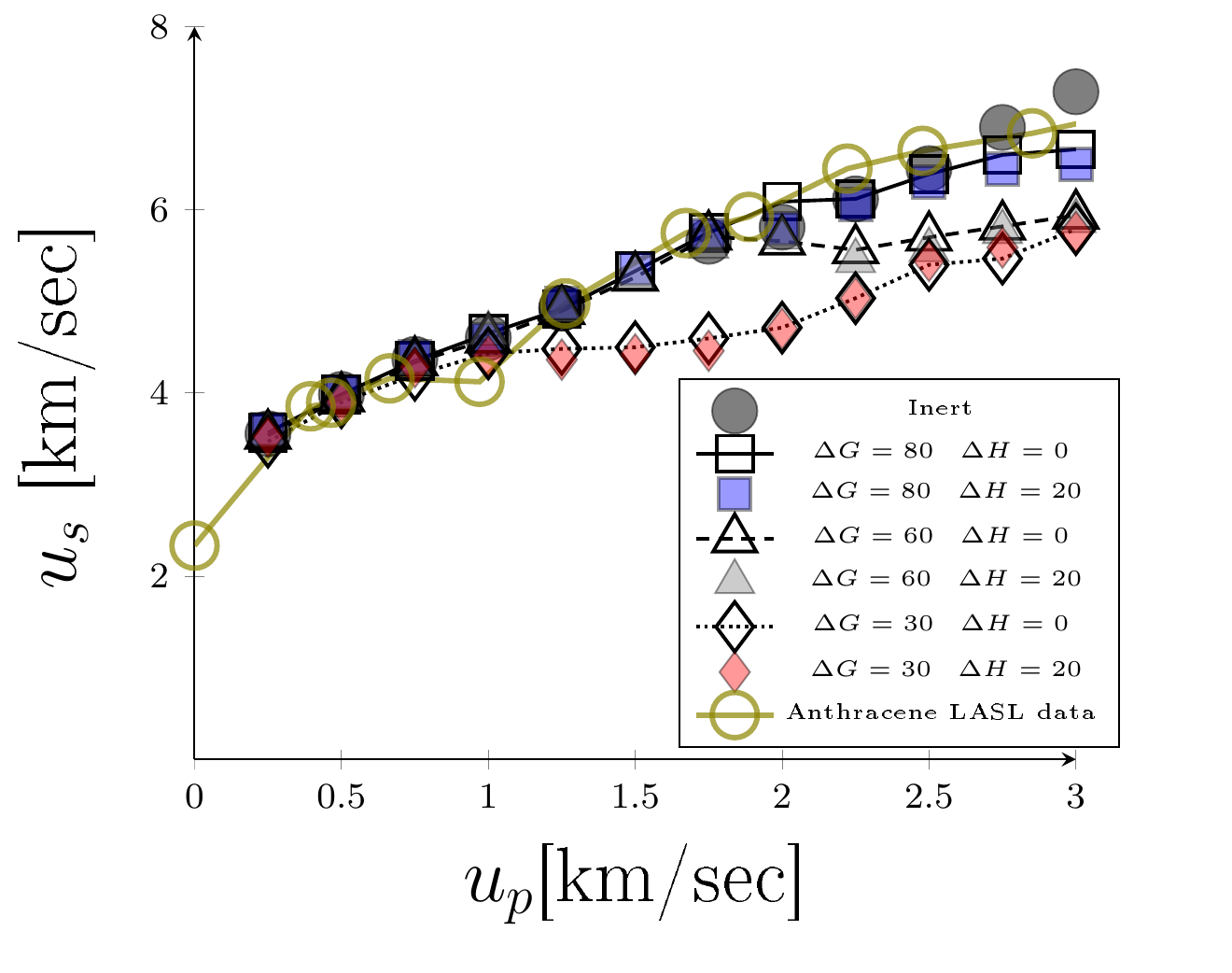}
  \end{tabular}
  \caption{Hugoniot and u$_s$-u$_p$ curves for an Inert and SWED materials with a curvature on the intramolecular potential corresponding to $(\Delta V/V_0)_{vdW} = 20$\%.  Different symbols correspond to different barriers. Empty symbols correspond no endothermicity ($\Delta H = 0$ kcal/mol) and filled symbols corresponds to $\Delta H = 20$ kcal/mol}
    \label{fig:Endo}
\end{figure}

The effect of endothermicity was found to be minimal for all the cases considered in this study. 
Figure \ref{fig:Endo} compares the effect of endothermicity on the Hugoniot and u$_s$-u$_p$ curves for an inert and SWED materials with a 
curvature on the intramolecular potential corresponding to $(\Delta V/V_0)_{vdW} = 20$ \%.  
We see that, for cases where volume collapse is small and larger activation barriers, endothermicity can slightly alleviate the shock speed, and therefore, its
resultant pressure. We note that while the anthracene data matches the inert $(\Delta V/V_0)_{vdW} = 35$ \% case, in the $(\Delta V/V_0)_{vdW} = 20$\% case, the Hugoniot appears to match a reactive case with activation barriers between $30 - 60 $ kcal/mol. 
%The role of how chemistry directly affects the different energy transferring mechanisms is discussed next.

\section{Role of kinetics and lack of local equilibrium in the nucleation of chemistry}\label{sec:last}

In the previous sections, we characterized the interplay between chemistry and the shock response for a
family of SWED materials. We now focus on the molecular-level mechanisms that control the nucleation
of volume-collapsing chemical reactions. We are interested, specifically, in understanding how the translational
energy of the shockwave is transferred to the modes responsible for chemistry, i.e. the breathing mode, and 
how this process influences the initiation of chemistry. To do so, we study how the excitation energy from the shockwave is 
partitioned between the various meso, radial and internal modes and explore whether the coupling rates between 
them can influence chemical reactions.

\begin{figure}[h!]
   \begin{tabular}{cc}
     \vspace{-.25cm}\hspace{-1.2cm}\includegraphics[scale=0.67]{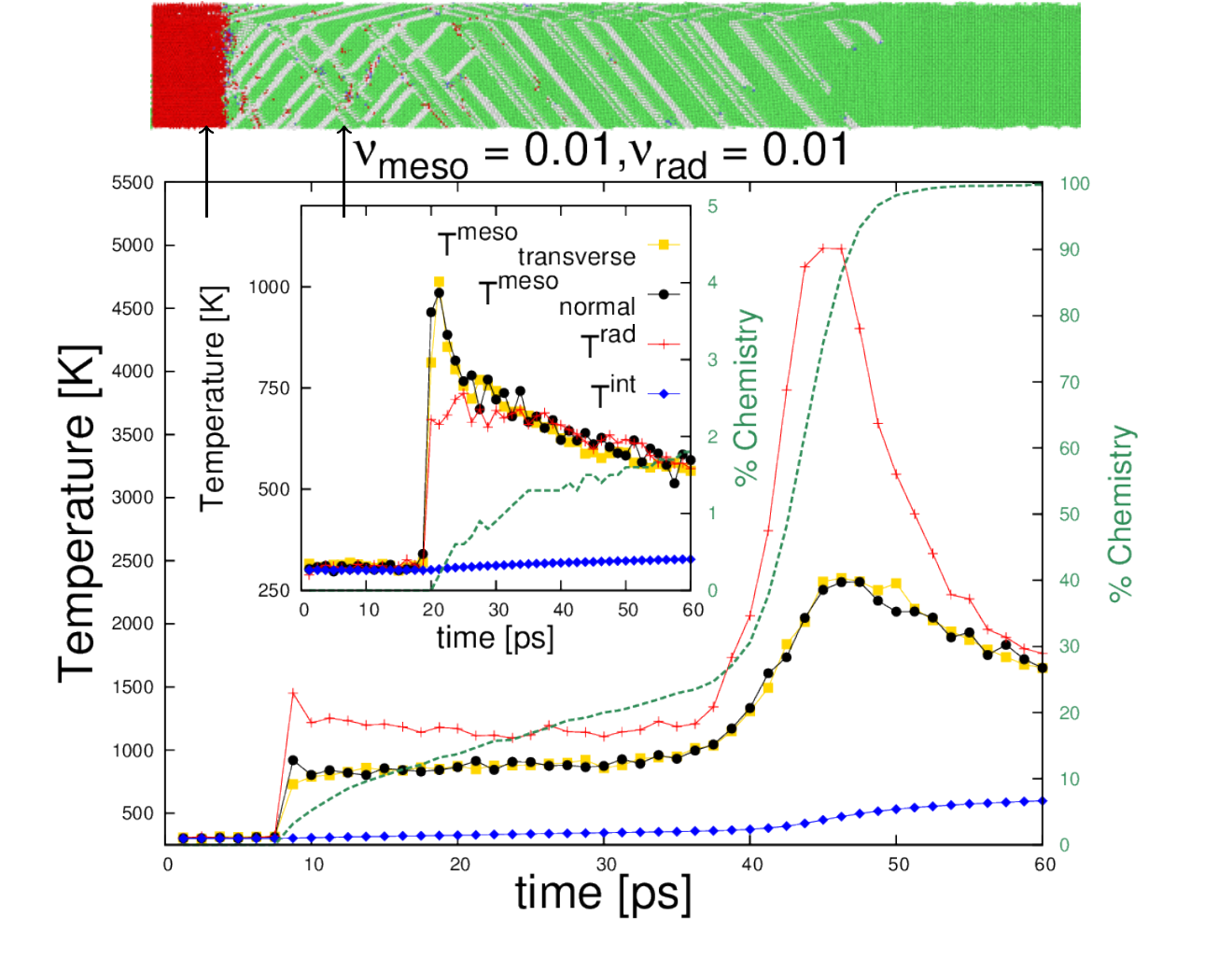} &  \hspace{-.9cm} \includegraphics[scale=0.67]{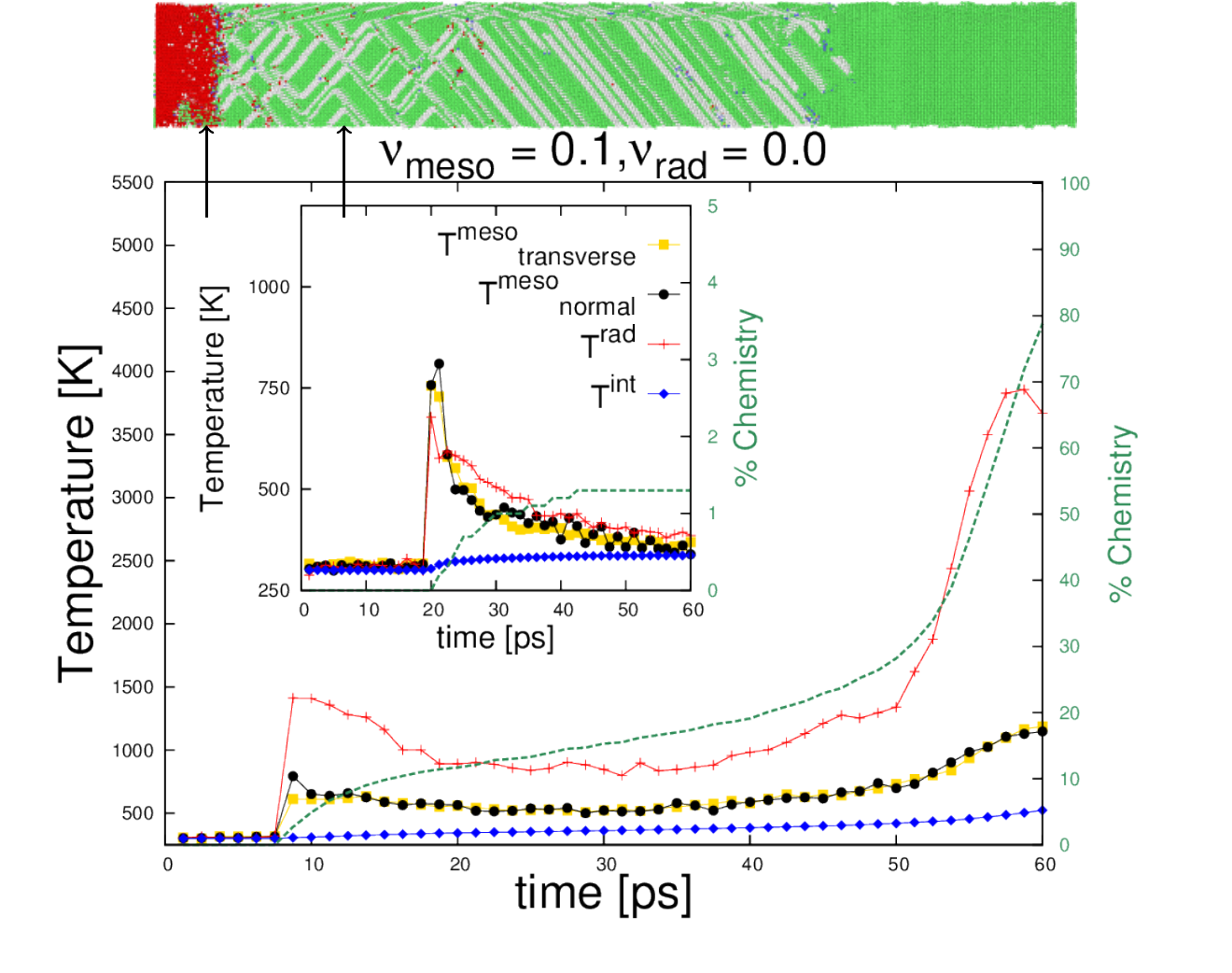} \\
      \vspace{.25cm}\hspace{-1.2cm} (A) &  \hspace{-1.2cm} (B)  \\
      \hspace{-1.2cm}\includegraphics[scale=0.67]{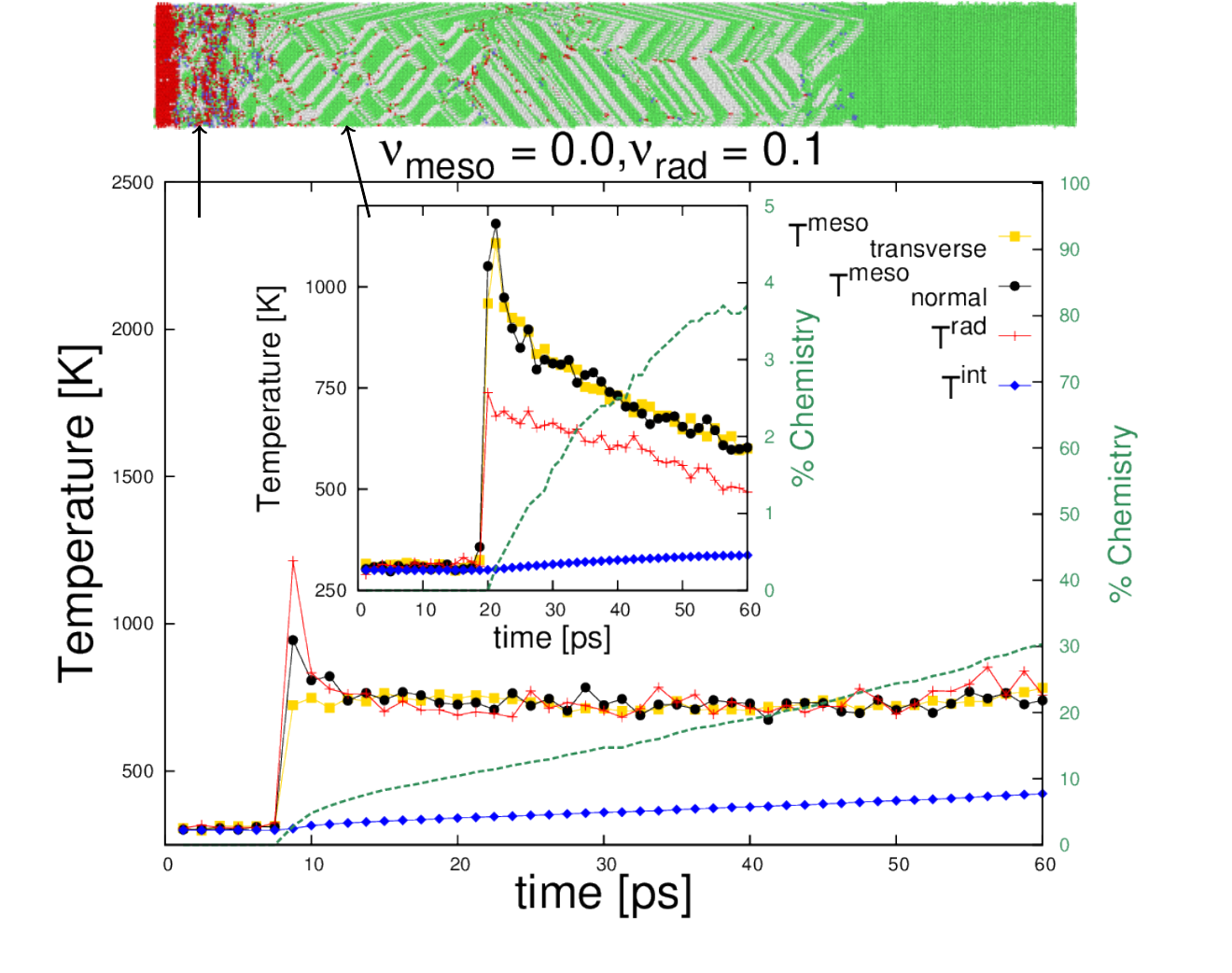} &  \hspace{-.9cm} \includegraphics[scale=0.67]{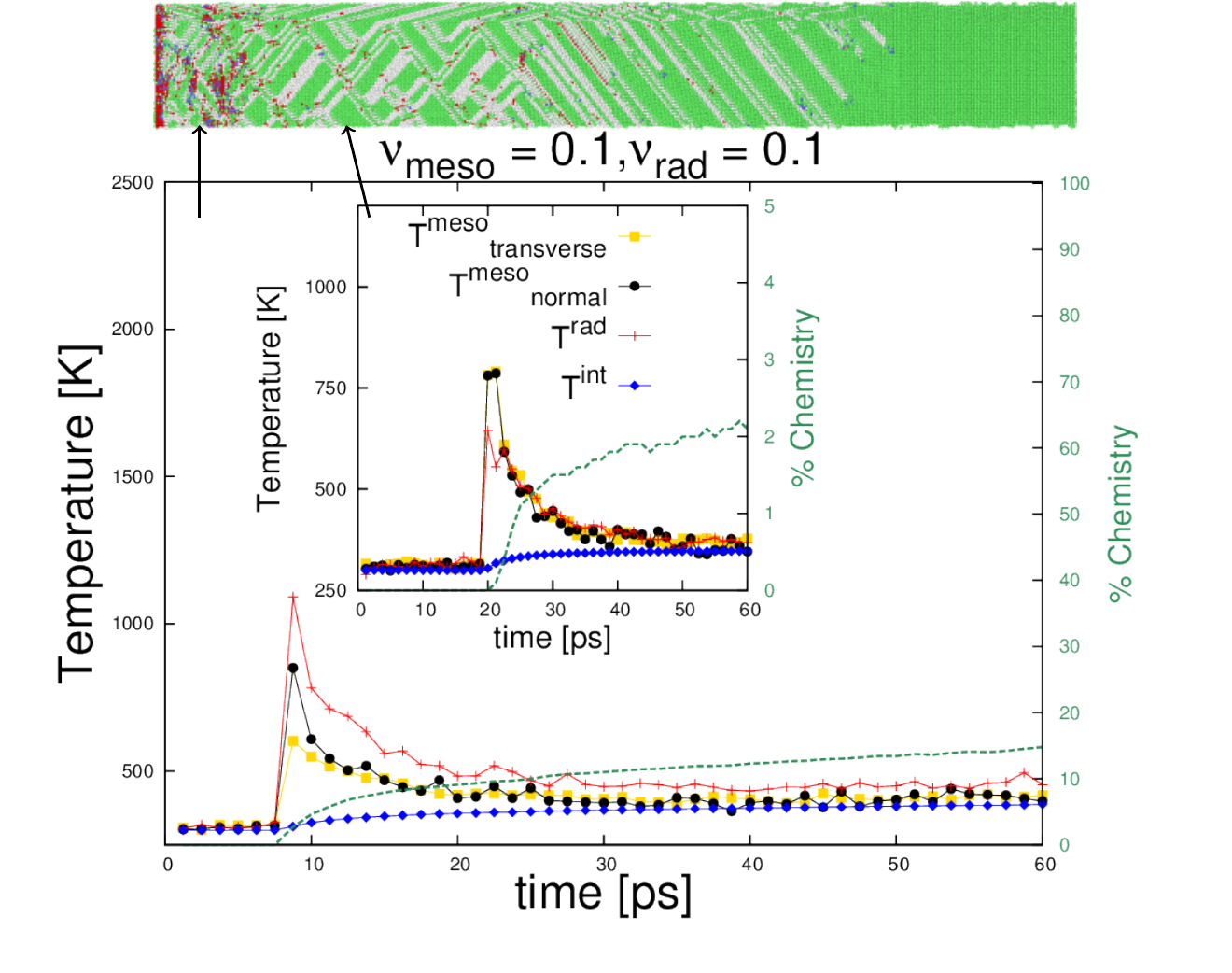} \\
      \hspace{-1.2cm} (C) &  \hspace{-1.2cm} (D) 
  \end{tabular}
  \caption{Atomic snapshots showing the chemical conversion (red particles) for various rates of the inter-molecular $\nu_{meso}$ and intra-molecular $\nu_{rad}$ couplings. The plots below follow the time dependence of the temperature components for two thin slabs at the positions denoted by the arrows. Varying the rate of inter-molecular and intra-molecular couplings quenches the amount of chemistry (dashed line) at different rates as shown on the scale on the right}
  \label{fig:TSequil}
\end{figure}

We focus on the response of a SWED material with $\Delta G = 30$ kcal/mol and $(\Delta V/V_0)_{vdW}$ = 35 \%.
Such a sample will be shocked at $u_p$ = 0.75 km/s as this corresponds to the HCL for this set of parameters.
Figure \ref{fig:TSequil} shows the time evolution of the temperatures associated with the three sets of DoFs
for two thin slabs of material; one 20 nm away from the impact surface and another 60 nm away (inset in figure).
The set of plots capture the effect of the coupling constants between the implicit DoFs and the particles c.m.
($\nu_{meso}$) and radial (or breathing) modes ($\nu_{rad}$). The meso temperature is further divided into normal 
(along the shock direction) and transverse modes. A dashed green line represent the percentage of 
reactive molecules as a function of time. Atomics snapshots of reacted particles and crystal structure are also shown 
at time 50 ps; fcc is colored in green, bcc in blue, hcp in white, and reacted particles in red. It is very clear that the
rates associated with energy transfer affect chemistry.

As the shock passes through the material, the molecular system will be away from local equilibrium with the modes that couple more 
strongly to the shock having higher temperatures.  
In all cases, the energy in the shockwave initially excites the particles c.m. and breathing modes as these modes are involved
directly in the propagation of the compression wave. These modes achieve high temperatures in very short timescales
and this excitation is transferred to the internal modes over longer timescales, which depend on the coupling rates, $\nu_{mess}$ and
$\nu_{rad}$. This is consistent with all atom MD simulation, see for example Refs.  \cite{JaramilloPRB2007,StrachanPRL2005}. 

The high temperature in the radial modes, together with the high pressure caused by the shockwave facilitates the
volume-reducing chemical reactions. As the barrier for chemical reactions is overcome and the molecules relax to the metastable,
low-volume state, the breathing modes are excited and cases with significant chemistry are marked by higher transient temperatures in
the breathing modes. On the other hand, in regions with little chemistry, the mesoscale and radial temperatures achieve equilibrium
faster. 

Interestingly, ChemDID predicts that chemical reactions can be initiated within short timescales, during this non-equilibrium stage. Thus, 
the timescales of energy exchange between the various sets of modes which determine the range of temperature experienced by the breathing 
modes play an important role in the chemical reactions. The nucleation of chemistry is also facilitated by high stresses and nucleation is
observed predominantly in active slip planes where plastic deformation is localized. This can be seen in the atomic snapshots shown in 
Fig. \ref{fig:TSequil}. 

\begin{figure}[h!]
  %\par\vspace{-10mm}
  \centering
  \begin{tabular}{c}
  \includegraphics[scale=0.75]{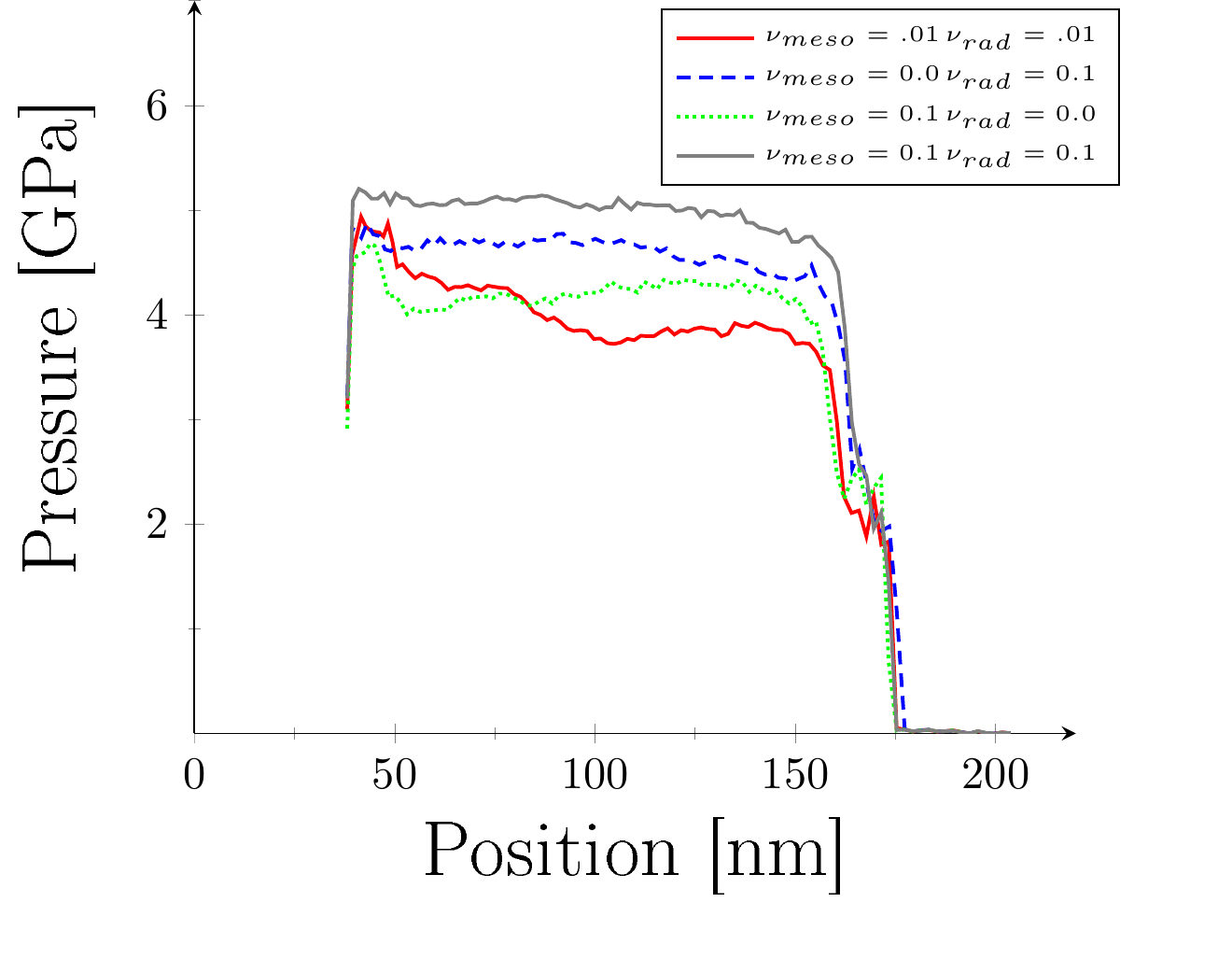} 
  \end{tabular}
  \caption{ Pressure profiles at time = 50 ps after impact for various coupling values}
  \label{fig:thresh_Profiles}
\end{figure}

We now focus on the effect of energy coupling rates on the chemical response of the materials. To help in the discussion, Fig. \ref{fig:thresh_Profiles} shows 
the pressure profiles for different coupling rates we have used in Fig.\ref{fig:TSequil}. For weak coupling rates to the internal DoFs, 
$\nu_{meso}=\nu_{rad}=0.01$ ps$^{-1}$ in Fig. \ref{fig:TSequil} (a), the radial and mesoscopic temperatures remain in the non-equilibrium, high-temperature
state for extended periods of time leading to significant chemical reactions. The propagation of a chemical front leads to a significant reduction in pressure behind the 
leading shockwave, see red line in Fig. \ref{fig:thresh_Profiles}. As a result, this pressure reduction leads to a significant quenching of chemistry on the region behind 
the shock, except between active slip planes with high stress.

Increasing the internal-to-intermolecular coupling ($\nu_{meso}=0.1$ ps$^{-1}$) and removing the internal-to-radial coupling ($\nu_{rad}=0.0$ ps$^{-1}$), leads to a 
reduction in chemical conversion throughout the sample. This occurs rather indirectly, as the internal temperature couples to the mesoscopic temperature via the
DID equations and the latter couples to the radial DoFs via the Hamiltonian. The result is that the radial temperatures are lower compared to the case 
with weak coupling discussed above, see Fig. \ref{fig:TSequil} (b). Now reversing the implicit couplings, i.e. $\nu_{rad}=1.0$ ps$^{-1}$ and $\nu_{meso}=0.0$ ps$^{-1}$,
leads to faster cooling on the radial temperature by the internal DoFs and a much slower chemical reaction front propagation. As expected, quenching the 
amount of chemistry near the impact leads to  higher pressure in the region behind the leading shock, and as a result, to slightly more chemistry (4 \%) in the region
behind the leading shock, compared to the other two previous cases (<2\%).

Lastly, strong coupling between internal and meso and radial modes, $\nu_{meso}=\nu_{rad}=0.1$ ps$^{-1}$, results in rapid equilibration
of the radial and mesoscale temperatures with the internal temperature. We see equilibration with all the temperatures occurring within 30 ps 
throughout the whole sample, leading to a fast quenching effect on chemistry production. The pressure for this case, is the highest compared 
to all the other cases. As a result we see a large fraction of nucleation of chemical reaction behind the leading shock, but interestingly, these 
nucleations do not subsequently grow into a propagating chemical wave. Our simulations clearly show that the ability of the chemical reactions 
to weaken shock waves is also affected by the details of the energy transfer between modes in the molecules.

\section{Conclusions}

In this paper we used ChemDID mesoscale simulations to explore the shock response of materials that can undergo endothermic,
volume reducing chemical reactions. The simulations demonstrate that such stress-induced chemical reactions can be effective at 
dissipating the shockwave and such materials could be valuable in applications requiring protection from dynamical loads.

We find that when a critical shock strength is reached, a chemical reaction wave is induced behind the leading shock wave. The chemical
reactions weaken the leading shockwave. For a range piston velocities above this hugoniot chemical limit we observe a three-wave structure,
with the elastic and plastic waves traveling at similar speeds and a the chemical wave trailing behind. Increasing the piston velocity results in 
an increase in the velocity of the chemical wave but does not affect the velocity leading shock nor the corresponding pressure. 
Under this regime, the material weakens the shock. This effect continues until the chemical wave moves at the same speed as the leading wave. 
In such an overdriven driven regime an increase in the shock velocity is translated directly to the leading shock.

The results of the simulations together with an analysis of the shockwave structure using conservation laws show that both the amount of volume 
collapse and the velocity of the chemical wave (governed by the activation energy associated with the chemical reactions) are the critical parameters 
for shockwave attenuation. The endothermicity of the chemical reactions plays a secondary role that is only discernible in the case of high barriers and 
modest volume collapse. Such information should be useful in the design and optimization of materials for shockwave energy dissipation.

Interestingly, energy transfer rates between the various modes in the material also play a role in the nucleation and propagation of chemical reactions.
Dynamical loading of materials leads to non-equilibrium states right behind the shockwave; with various DoFs experiencing
different temperatures. Chemical reactions are influenced by the temperature history of the corresponding modes 
and the coupling rates between them. Recent reactive MD simulations of solid explosives also indicate the possibility of chemical reactions
away from local equilibrium following shock loading.\cite{Cherukara2014} The interplay between plasticity and chemistry will also affect resultant chemical reactions
and shock attenuation in the material. Whereas our simulations presented focused on single crystalline materials, 
ChemDID could also be a useful tool to explore the effect of microstructure and defects on shock induced chemistry. 
This would require large-scale simulations  and we are working on an implementation 
of our mesoscale model in the LAMMPS parallel simulator that would enable such studies.

For a given piston speed, a specific volume collapse provides a limited range where chemistry is able alleviate pressure buildup.
Higher volume-collapse reduces the shock pressure the most, but also reduces the shock velocity corresponding to the aforementioned 
overdriven state. Avoiding this overdriven regime may be of interest for some applications. As a function of piston speeds, the larger activation 
energies can postpone the onset of chemistry to higher piston speeds before it reaches the overdriven state. 
Therefore, it is important to take both of these effects into consideration in order to design SWED materials that achieve the maximum 
amount of pressured dissipated over a desired range of velocities.

In this paper we focused on the idealized case of sustained shocks and model materials. Additional complexity is introduced when considering
finite pistons that lead to reflection and rarefaction waves that depend on the relative size of the target to impactor. In such cases non-steady states
need to be considered, yet the general conclusions presented in this paper are useful for such cases.
Finally, ChemDID can be parameterized to describe specific materials of interest. This would likely require more complex inter-molecular
interaction potentials and chemical kinetics matching the real case. While developing such models is challenging, the technique would enable
achieving time and length scales well beyond what is possible today with all-atom MD.

\newpage

\section{appendix}
\subsection{Pressure quantification}\label{sec:apdxB}
We write an expression for the pressure between a chemical region \{c\} and a shocked region \{s\} where none of the components are at rest. 
The transition across this region has to satisfy momentum conservation:
\begin{align}
  P_c + \rho_c (\dot{x}_c - u_c)^2  & =  P_s + \rho_s (\dot{x}_s - u_c)^2  
\end{align}
Let us expand this expression, we get 
\begin{align}
  P_c - P_s & = \rho_s (\dot{x}_s - u_c)^2  -\rho_c (\dot{x}_c - u_c)^2   \nonumber \\
  & = \rho_s (\dot{x}_s^2 + u_c^2 - 2  \dot{x}_s u_c ) - \rho_c (\dot{x}_c^2 + u_c^2 - 2  \dot{x}_c u_c )     \nonumber \\
  & = \rho_s \dot{x}_s^2   -\rho_c \dot{x}_c^2  + u_c [ \rho_s (u_c - 2 \dot{x}_s) -\rho_c (u_c - 2 \dot{x}_c)]  \nonumber \\
  & = \rho_s \dot{x}_s^2 - \rho_c \dot{x}_c^2 + u_c [ \rho_s (u_c - \dot{x}_s) -\rho_c (u_c - \dot{x}_c) +  \rho_c \dot{x}_c  - \rho_s \dot{x}_s] \nonumber \\ 
  & = \rho_s \dot{x}_s^2 - \rho_c \dot{x}_c^2 + u_c [ \rho_c \dot{x}_c - \rho_s \dot{x}_s]\nonumber \\
  & = \rho_c \dot{x}_c (u_c - \dot{x}_c)  + \rho_s \dot{x}_s (\dot{x}_s-u_c) \nonumber 
\end{align} 
,where we have made use of mass conservation ( Eqn, \ref{eqn:vc} ) above. Substituting Eqn. \ref{eqn:xsdot}, twice, and rearranging, we obtain the final form
\begin{align}
  P_c - P_s  & =   (u_c - \dot{x}_c)^2 \rho_c (\frac{\rho_c}{\rho_s} - 1 )
\end{align}

\bibliography{arXiv}{}

\end{document}